\documentclass[12pt]{article}
\usepackage{amsmath,amssymb,amsfonts,amsthm,amscd}
\usepackage[english]{babel}
\usepackage[toc]{appendix}
\usepackage{graphicx}
\usepackage{float}
\newcommand{\be}{\begin{equation}}
\newcommand{\ee}{\end{equation}}
\newcommand{\bea}{\begin{eqnarray}}
\newcommand{\eea}{\end{eqnarray}}

\newcommand{\p}[1]{(\ref{#1})}

\def\theequation{\arabic{section}.\arabic{equation}}
\begin{document}
\begin{titlepage}
\vspace*{0.7cm}

\begin{center}
\baselineskip=16pt
{\LARGE\bf Multiparticle $\mathcal{N}{=}\,8$ mechanics }

\vspace{0.3cm}

{\LARGE\bf with $F(4)$ superconformal symmetry}

\vspace{1.5cm}

{\large\bf Sergey\,Fedoruk,} \quad {\large\bf Evgeny\,Ivanov}
\vspace{0.8cm}

{\it Bogoliubov Laboratory of Theoretical Physics, JINR,\\
Joliot-Curie 6, 141980 Dubna, Moscow region, Russia} \\
\vspace{0.5cm}

{\tt fedoruk,eivanov@theor.jinr.ru}

\end{center}

\vspace{2.5cm}

\par
\begin{center}
{\bf Abstract}
\end{center}
We present a new multiparticle model of $\mathcal{N}{=}\,8$ mechanics with superconformal $F(4)$ symmetry.
The system is constructed in terms of two matrix $\mathcal{N}{=}\,4$ multiplets. One of them is a bosonic
matrix $({\bf 1, 4, 3})$ multiplet and another is a fermionic $({\bf 0, 4, 4})$ one. Off-diagonal bosonic components
of the $({\bf 1, 4, 3})$ multiplet are chosen  to take values in the flag manifold  $\mathrm{U}(n)/[\mathrm{U}(1)]^n$
and they carry additional gauge symmetries. The explicit form of the $F(4)$ supersymmetry generators is found.
We demonstrate that the $F(4)$ superalgebra constructed contains as subalgebras two different
$D(2,1;\alpha\,{=}{-}1/3)$ superalgebras intersecting over the common $sl(2,\mathbb{R})\oplus su(2)$ subalgebra.


\vspace{2.5cm}



\end{titlepage}


\numberwithin{equation}{section}

\section{Introduction}
\quad\, The models of superconformal mechanics occupy a notable place in the study of the AdS/CFT correspondence
in supersymmetric gauge theories. This is basically due to the fact that the one-dimensional conformal $\mathrm{SL}(2,\mathbb{R})$ symmetry
naturally emerges as a symmetry of the near horizon geometries of the  black-hole solutions of the appropriate supergravities.

The superconformal mechanics systems, pioneered in eighties in the papers \cite{AP,FR,IKL2},
have been so far worked out mainly up to the case of ${\cal N}{=}\,4$, $d\,{=}\,1$ extended supersymmetry
(see, e.g., refs.\,\cite{7,nscm,IL,ikl,IvKrN,ikl1} and the review \cite{superc}).
The models with ${\cal N}{=}\,8$ superconformal symmetry were studied to much less  extent
\cite{BIKL,F4scm,AzKuTo,KLS-18}.\footnote{
In \cite{KuTo,KhoTo} the ${\cal N}{=}\,8$ superconformal algebras  were derived in
terms of the so called $D$-module representations.}
But it is just ${\cal N}{=}8$ superconformal mechanics which is most important from the standpoint of the AdS/CFT correspondence
(see a recent review \cite{F4-r} and references therein). Moreover, the important role in this context is played  by the exceptional
${\cal N}{=}8$ superconformal symmetry $F(4)$
(see, for example, recent papers \cite{F4-n,F4-l}).\footnote{One of the first applications
of the exceptional superalgebra $F(4)$ comes back to refs.\,\cite{FrLin-1,FrLin-2}.}

The first example of ${\cal N}{=}\,8$ superconformal mechanics with $F(4)$ supersymmetry was presented in \cite{F4scm}.
The one-particle system considered there was underlain by an interaction of two ${\cal N}{=}\,4$ multiplets:
the bosonic $({\bf 1, 4, 3})$ and the fermionic $({\bf 0, 4, 4})$ ones. In the present paper we consider a matrix generalization
of this system and, as an outcome, obtain a new model of the multiparticle ${\cal N}{=}\,8$  superconformal mechanics.

The matrix models are an efficient tool of constructing conformally invariant systems \cite{Poly-gauge,Poly2001,Poly-rev}.
In particular, it was found in \cite{FIL,FI,FILS} that the matrix one-dimensional superfield models
yield Calogero-like systems with ${\cal N}{=}\,4$ supersymmetries after exploiting the appropriate gauging procedure \cite{2}.
The physical bosonic degrees of freedom were described by the diagonal elements of dynamical bosonic matrix of
the $({\bf 1, 4, 3})$ multiplets. The off-diagonal entries of this matrix proved to represent the purely gauge degrees of freedom.

As opposed to the gauging approach of refs.\,\cite{FIL,FI,FILS}, in this paper all bosonic variables including the off-diagonal components
of the $({\bf 1, 4, 3})$ matrix multiplet are treated as dynamical.
These off-diagonal fields parametrize the target space of flags
${\displaystyle \frac{\mathrm{U}(n)}{\mathrm{U}_1(1)\otimes\ldots\otimes \mathrm{U}_n(1)}}$. So,
they can be interpreted as a kind of the target harmonics, while the corresponding part of the worldline action as that
of supersymmetric $d\,{=}\,1$ sigma model on such a manifold.\footnote{The two-dimensional bosonic flag-manifold sigma models were studied,
e.g., in \cite{Bykov,TS-2018,OhSeSh}.
The $\mathrm{SU}(3)/[\mathrm{U}(1){\otimes}\mathrm{U}(1)]$ harmonics play the crucial role in  the off-shell
formulation of ${\cal N}{=}\,3$, $d\,{=}\,4$ super Yang-Mulls theory~\cite{GIKOS-N3}.}

The plan of the paper is as follows. In Section 2 we present ${\cal N}{=}\,4$ harmonic superfield description of
the matrix multiplets $({\bf 1, 4, 3})$ and  $({\bf 0, 4, 4})$. Also we introduce ${\cal N}{=}\,8$ superconformally invariant interaction
of these multiplet and find, by Noether procedure, the supercharges generating the relevant ${\cal N}{=}\,8$ conformal superalgebra.
In Section 3 we split the matrix $n^2$ bosonic fields into the sets of $n$ diagonal (radial) and $n^2-n$ non-diagonal (angular) ones.
The latter fields are identified with the target
${\displaystyle \frac{\mathrm{U}(n)}{\mathrm{U}_1(1)\otimes\ldots\otimes \mathrm{U}_n(1)}}$ harmonics
on which some additional gauge symmetries are realized.
In Section 4 we eliminate auxiliary fields and pass to the physical variables.
Then we fulfill the Hamiltonian analysis of the system.
We find the corresponding Hamiltonian and the relevant set of constraints. With respect to the second-class constraints,
we introduce Dirac brackets. In  Section 5 we show that ${\cal N}{=}\,8$ conformal superalgebra of our system
is just $F(4)$ with $so(7)$ as R-symmetry algebra  and write down the explicit form of ${\cal N}{=}\,8$ supercharges.
We demonstrate  that the underlying  $F(4)$ superalgebra contains two $D(2,1;\alpha{=}{-}1/3)$ superalgebras
with the common $sl(2,\mathbb{R})$ and $su(2)$ subalgebras.\footnote{An analogous closure property for some other $d\,{=}\,1$ superconformal algebras
was observed in \cite{IST,ILS1,ILS2}.} Some concluding remarks are collected in the last Section 6.

\section{Superconformal coupling of the matrix multiplets $({\bf 1, 4, 3})$ and  $({\bf 0, 4, 4})$}

\quad\, The powerful approach to constructing ${\cal N}{=}\,4$, $d\,{=}\,1$ supersymmetric models and finding interrelations between them
is ${\cal N}{=}\,4$, $d\,{=}\,1$ harmonic formalism which was proposed in \cite{IL}.
As compared to the description in the usual superspace with the coordinates $z=( t, \theta_i,\bar\theta^i)$, $({\theta_i})^* = \bar\theta^i$
and covariant derivatives
\be
D^i = \frac{\partial}{\partial \theta_i} -
i\bar\theta^i \partial_t\,, \quad \bar D_i = \frac{\partial}{\partial \bar\theta^i} - i\theta_i \partial_t\,, \qquad (D^i)^* = -\bar D_i\,,
\qquad \{D^i, \bar D_k \} = -2i\,\delta^i_k\partial_t\,, \label{defD2}
\ee
the harmonic description involves additional commuting harmonic variables
\be\label{h-2-st}
u^\pm_i\,, \quad (u^+_i)^* = u^-{}^i\,,\qquad u^{+
i}u_i^- =1\,.
\ee

In the harmonic analytic basis
\be\label{h-space}
z_A = ( t_A, \theta^\pm, \bar\theta^\pm, u^\pm_i)\,,\qquad t_A = t +i (\theta^+\bar\theta^- + \theta^-\bar\theta^+), \quad
\theta^\pm =\theta^iu^\pm_i\,, \quad  \bar\theta^\pm = \bar\theta^iu^\pm_i
\ee
half of the ${\cal N}{=}4$ covariant spinor derivatives
$D^\pm = u^\pm_i D^i$, $\bar D^\pm = u^\pm_i \bar D^i$  becomes short:
\be
D^+ = \frac{\partial}{\partial \theta{}^-}\,, \qquad \bar D^+ = -\frac{\partial}{\partial \bar\theta{}^-} \,.\label{Short}
\ee
This implies the existence of the harmonic analytic superfields defined on the analytic subspace of the full harmonic superspace:
\be
(\zeta, u) = (t_A, \theta^+, \bar\theta^+, u^\pm_i)\,, \quad u^{+
i}u_i^- =1\,.
\ee
It is closed under both ${\cal N}{=}\,4$ supersymmetry and ${\cal N}{=}\,4$ superconformal symmetry.
The integration measure in the harmonic analytic subspace is defined as
$dud\zeta^{(-2)} =dudt_{A}d\theta^+d\bar\theta^+$.
An important tool of the formalism is the harmonic derivatives:
\be
D^{\pm\pm}=\partial^{\pm\pm}+2i\theta^\pm\bar\theta^\pm\partial_{t_{A}} +\theta^\pm\frac{\partial}{\partial\theta^\mp} +
\bar\theta^\pm\frac{\partial}{\partial\bar\theta^\mp}\,,  \qquad \partial^{\pm\pm} =
u^{\pm}_i\frac{\partial}{\partial u^{\mp}_i}\,.
\ee
The harmonic derivative $D^{++}$ is distinguished in that it commutes with the spinor derivatives \p{Short} and so preserves the analyticity.

Here we presented only the definitions of the basic notions which will be used below.
The full description of the harmonic superspace approach to $d\,{=}\,1$ models is given in ref. \cite{IL}
(details of the harmonic formulation of the multiplets which will be considered in this paper can be found in \cite{superc,FI-2015}).

\subsection{Fermionic matrix multiplet $({\bf 0, 4, 4})$}

\quad\, The multiplet $({\bf 0, 4, 4})$ is a fermionic analog of the multiplet $({\bf 4, 4, 0})$.
We consider $n^2$ multiplets $({\bf 0, 4, 4})$ described off shell by the fermionic $n{\times}n$ matrix
analytic superfield $\Psi^{+ A}:=\|\Psi^{+ A}{}_a{}^b\|$\,,
$\widetilde{(\Psi^{+ A})} = \Psi_A^+$\,, $A=1,2$\,, $a=1,\ldots,n$\,, which satisfies the constraint \cite{IL}:
\be
D^{++}\Psi^{+ A} = 0\qquad \Rightarrow \qquad \Psi^{+ A} =
\phi^{iA}u^+_i + \theta^+ F^A + \bar\theta^+ \bar{F}^A
- 2i\theta^+\bar\theta^+ \dot{\phi}{}^{i A}u^-_i\,,  \label{PsiConstr}
\ee
where $({\phi^{iA}})^\dagger=-\phi_{iA}$, $({F^A})^\dagger=\bar{F}_A$ or, in terms of the matrix entries,
$$
({\phi^{iA}}_a{}^b)^*{=}\,-\phi_{iA}{}_b{}^a\,,\qquad
({F^A}_a{}^b)^*{=}\,\bar{F}_A{}_b{}^a \,.
$$
On the doublet index $A\,{=}\,1,2$ the appropriate $SU(2)_{PG}$ group acts.
It commutes with the ${\cal N}{=}\,4$ superconformal
$D(2,1;\alpha)$ transformations which are realized on the component fields as (see, e.g., \cite{FI-2015}):
\be\label{tr-comp-044}
\begin{array}{c}
\delta\phi^{iA}=-\left( \omega^i F^A+\bar\omega^i \bar F^A\right), \\ [6pt]
\delta F^A=2i\,\bar\omega^k\dot\phi^A_k +2i\alpha\,\bar\eta^k\phi^A_k\,,\qquad
\delta \bar F_A=2i\,\omega_k\dot{\phi}^{k}_A +2i\alpha\,\eta_k {\phi}^{k}_A\,.
\end{array}
\ee
Here,
$$
\omega_i= \varepsilon_i -t\,\eta_i\,,\qquad
\bar\omega^i= \bar\varepsilon^i - t\,\bar\eta^i\,,
$$
and $\bar\varepsilon^i=\overline{(\varepsilon_i)}$, $\bar\eta^i=\overline{(\eta_i)}\,$.

In the central basis, the constraint \p{PsiConstr} and the analyticity conditions $D^+\Psi^{+ A}{=} \bar D^+\Psi^{+ A}{=}\,0$
imply
\be
\Psi^{+ A}(z, u) = \Psi^{i A}(z)u^+_i\,,  \qquad
D^{(i} \Psi^{k) A}(z) = \bar D^{(i} \Psi^{k) A}(z) = 0\,.\label{ConstrPsi2}
\ee

The free action of the matrix superfield $\Psi^{+ A}$,
${\displaystyle S^{(\Psi)}_{free} \sim \mathrm{Tr} \int du d\zeta^{(-2)}\,\Psi^{+ A}\Psi^{+}_{A}}$,
is not invariant under ${\cal N}{=}\,4$, $d\,{=}\,1$ superconformal group $D(2,1;\alpha)$, except for the special case of
$\alpha\,{=}\,0\,$, which will be of no interest for us here. As we will see, the superconformal versions of the free $\Psi^{+ A}$ action,
which are valid for any $\alpha$, can be constructed by means of coupling this multiplet to those considered in the next sections.

\subsection{Bosonic matrix multiplet $({\bf 1, 4, 3})$}

\quad\, The off-shell $n^2$ multiplets ${\bf (1, 4, 3)}$ are described by
an Hermitian ${\cal N}{=}\,4$  matrix superfield $\mathcal{M}(z):=\|\mathcal{M}_a{}^b(z)\| $\,, $\widetilde{\mathcal{M}} = \mathcal{M}$\,,
obeying the constraints \cite{leva}
\be
D^iD_i \mathcal{M} = \bar D_i\bar D^i \mathcal{M} = 0\,, \quad [D^i, \bar D_i] \mathcal{M} = 0\,. \label{Uconstr1}
\ee
These constraints are solved by
\begin{equation}  \label{sing-X0-WZ}
\mathcal{M}(t,\theta_i,\bar\theta^i)= M + \theta_i\varphi^i +
\bar\varphi_i\bar\theta^i +
i\theta^i\bar\theta^k A_{ik}-{\textstyle\frac{i}{2}}(\theta)^2\dot{\varphi}_i\bar\theta^i
-{\textstyle\frac{i}{2}}(\bar\theta)^2\theta_i\dot{\bar\varphi}{}^i +
{\textstyle\frac{1}{4}}(\theta)^2(\bar\theta)^2 \ddot{\mathcal{M}}\,,
\end{equation}
where $(\theta)^2=\theta_k \theta^k$, $(\bar\theta)^2=\bar\theta^k \bar\theta_k$ and
$M^\dag{=}M$, $({\varphi^i})^\dagger=\bar\varphi_i$, $({A^{ik}})^\dagger=A_{ik}=A_{(ik)}$
or, in the more detailed notation,
$$
(M_a{}^b)^*{=}M_b{}^a\,,\qquad ({\varphi^i}_a{}^b)^*{=}\,\bar\varphi_i{}_b{}^a\,,\qquad
({A^{ik}}_a{}^b)^*{=}\,A_{ik}{}_b{}^a\,.
$$

The same constraints \p{Uconstr1}, being rewritten in the harmonic superspace, read
\be
D^{++}\mathcal{M} =0\,, \qquad D^+D^-\mathcal{M} = \bar D^+\bar D^-\mathcal{M} = 0\,, \qquad
\left(D^+\bar D^- + \bar D^+D^-\right)\mathcal{M} = 0\,.\label{Uconstr2}
\ee
The extra harmonic constraint guarantees the harmonic independence of
$\mathcal{M}$ in the central basis.

As was shown in \cite{2}, the $({\bf 1, 4, 3})$ multiplet can be also described in terms
of the real analytic gauge superfield prepotential. In the matrix case, we can introduce the matrix analytic superfield ${\cal V}(\zeta, u)$
defined up to the abelian gauge transformations
\be
{\cal V} \;\;\Rightarrow \;\; {\cal V}{\,}' = {\cal V} + D^{++}\Lambda^{--}\,, \quad
\Lambda^{--} = \Lambda^{--}(\zeta, u)\,.\label{VgaugeT}
\ee
In the Wess-Zumino gauge, just the irreducible ${\bf (1, 4, 3)}$ content is recovered
\be
{\cal V}_{WZ}(\zeta, u) = M(t_A) -2 \theta^+\varphi^i(t_A) u^-_i
-2 \bar\theta^+ \bar\varphi^i(t_A) u^-_i +
3i \theta^+\bar\theta^+ A^{(ik)}(t_A)u^-_iu^-_k\,. \label{WZV1}
\ee
The original matrix superfield $\mathcal{M}(z)$ is
related to ${\cal V}(\zeta, u)$ by the transform
\be
\mathcal{M}(t, \theta^i, \bar\theta_k)=
\int du\, {\cal V}\left(t +2i\theta^i\bar\theta^k u^+_{(i}u^-_{k)}\,,\, \theta^iu^+_i,
\bar\theta^ku^+_k\,,\, u^\pm_l\right). \label{DefU2}
\ee
The constraints \p{Uconstr1} emerge as a consequence of the harmonic
analyticity of  ${\cal V}\,$,
\be
D^+{\cal V} = \bar D^+{\cal V} = 0\,.\label{Vanalit}
\ee

The transformation properties of the component fields in the expansion \p{sing-X0-WZ} under
${\cal N}{=}\,4$ conformal supersymmetry $D(2,1; \alpha)$ are given by
\begin{equation}\label{tr-comp-143}
\begin{array}{c}
\delta M=-\omega_i\varphi^i+ \bar\omega^i\bar\varphi_i\,, \\ [6pt]
\delta \varphi^i=i \bar\omega^i\dot M-i \bar\omega_k A^{ki}-2i\alpha\,\bar\eta^i M  \,,
\qquad
\delta \bar\varphi_i=-i \omega_i\dot M -i \omega^k A_{ki} +2i\alpha\,\eta_i M
\,,\\ [6pt]
\delta A_{ik}=-2\left( \omega_{(i}\dot\varphi_{k)} +\bar\omega_{(i} \dot{\bar\varphi}_{k)}\right)
+2(1+2\alpha)\left( \eta_{(i}\varphi_{k)} +\bar\eta_{(i} {\bar\varphi}_{k)}\right)\,.
\end{array}
\end{equation}

The general $\varepsilon_i, \bar\varepsilon_k$-invariant superfield action of the multiplet ${\bf (1, 4, 3)}$ can be written as
\be
S_{gen}^{(\mathcal{M})} = \int dtd^4\theta\, {\cal L}_{gen}(\mathcal{M})\,.\label{genact_v}
\ee
The $D(2,1; \alpha)$ invariant action  (excepting the values of $\alpha\,{=}\,0,-1$)
is as follows \cite{ikl1}
\be \label{conf_v}
S_{sc}^{(\mathcal{M})} \ \sim \ \int dtd^4\theta \ \mathrm{Tr} \left(\mathcal{M}^{-1/\alpha}\right).
\ee
In this paper we will deal with the choice $\alpha\,{=}-1/3$. The $D(2,1; \alpha\,{=}-1/3)$ superconformally invariant action
for the ${\bf (1, 4, 3)}$ multiplet is so given by
\bea \label{conf_v-13}
S^{(\mathcal{M})(\alpha =-1/3)}_{sc} = - \frac{1}{6} \int dt d^4\theta\, \mathrm{Tr}\left(\mathcal{M}^3\right).
\eea
In the component notation,  the action \p{conf_v-13} takes the form
\begin{equation}\label{4N-X-WZ}
S^{(X)}_{sc} = \displaystyle{\int} dt\,
\mathrm{Tr}\left( M\dot M\dot M -\frac{i}{2}\,\{M, \bar\varphi^k\} \dot\varphi_k
-\frac{i}{2} \,\{M,\varphi_k\} \dot{\bar\varphi}^k
+ \frac{1}{2}\, M A^{ik}A_{ik}\, +  \frac{i}{2} \,A^{ik} [\varphi_{i},\bar\varphi_{k}] \right) .
\end{equation}
Using the transformations \p{tr-comp-143} with $\alpha\,{=}-1/3$,
it is easy to directly check $D(2,1; \alpha\,{=}-1/3)$ invariance of the action \p{4N-X-WZ}.

\subsection{Superconformal coupling}

\quad\, Proceeding from the  description of the multiplet $({\bf 1, 4, 3})$ through the analytic prepotential ${\cal V}$,  it is
easy to construct its superconformal coupling to $\Psi^{+ A}$ \cite{2}
\be \label{Confpsi}
S^{(\mathcal{M},\Psi)}_{sc} =\frac12 \int du d\zeta^{(-2)}\,\mathrm{Tr}\left({\cal V}\, \Psi^{+ A}\Psi^{+}_{ A}\right) .
\ee
This action is superconformal at any $\alpha\,{\neq}\,0\,$ and it also respects
the gauge invariance
\p{VgaugeT} as a consequence of the constraint \p{PsiConstr}.
An analysis based on dimensionality and the Grassmann character of the superfields
$\Psi^{+ A}, \Psi^{- A} = D^{--}\Psi^{+ A}$ shows that the coupling
\p{Confpsi} is the only possible coupling of this fermionic multiplet to the
multiplet ${\bf (1,4,3)}$, such that it  preserves the canonical number of time derivatives
in the component action (no more than two for bosons and no more than one for fermions).

It is easy to find the component-field representation of \p{Confpsi}
\bea \label{coup-143-comp}
S^{(\mathcal{M},\Psi)}_{sc} & = & \frac12 \int dt\, \mathrm{Tr}\Big( M\{F^A,\bar F_A\}-i\{M,\phi^{iA}\}\dot{\phi}_{iA}   \nonumber \\
&& \qquad \qquad \qquad  + \ iA^{ik}\,\phi^A_i\phi_{kA}  +\bar\varphi{}^k \{\phi_{kA}, F^A\}
- \varphi_k \{\phi^{kA},\bar F_A \} \Big) .
\eea

The total action is the sum of the superconformal ${\bf (1,4,3)}$ action \p{conf_v-13} and the action \p{Confpsi} describing the superconformal coupling ${\bf (1,4,3)}+{\bf (0,4,4)}$ (or the sum of the component actions \p{4N-X-WZ} and \p{coup-143-comp}):
\be \label{act-total}
S_{(\mathcal{M}+\Psi)}=S^{(\mathcal{M})}_{sc} + S^{(\mathcal{M},\Psi)}_{sc}  .
\ee

The variation of the total component action \p{act-total}
with respect to the transformations \p{tr-comp-044}, \p{tr-comp-143} can be represented as the integral
\be \label{var-c4-act}
\delta_\omega S_{(\mathcal{M}+\Psi)}=\int dt\, \dot\Lambda_\omega\,,
\ee
where
\begin{eqnarray}
\!\!\!\!\!\!\!\Lambda_\omega\!\! &\!=\!& -\frac12\,\omega_{i}\,\mathrm{Tr}\Big( \{M,\dot M\}\varphi^{i}  + \{M,\varphi_{k}\}A^{ik} -
i \bar \varphi^i\varphi^k \varphi_k + i \{M,\phi^{iA}\}F_{A} +2i\varphi_k\phi^{(iA}\phi^{k)}_A \Big)
\nonumber \\
\!\!\!\!\!&& +\frac12\,\bar\omega^{i}\,\mathrm{Tr}\Big( \{M,\dot M\}\bar\varphi_{i}  - \{M,\bar\varphi^{k}\}A_{ik} +
i \varphi_i\bar\varphi^k \bar\varphi_k - i \{M,\phi_{iA}\}\bar F^{A} -2i\bar\varphi^k\phi_{(i}^{A}\phi_{k)A} \Big)
\nonumber \\
\!\!\!\!\!&& +\frac23\,\eta_{i}\,\mathrm{Tr}\Big( M^2\varphi^{i} \Big)-
\frac23\,\bar\eta^{i}\,\mathrm{Tr}\Big( M^2\bar\varphi_{i} \Big)\,.
\label{Lambda-4}
\end{eqnarray}
Using this property, the Noether $\mathcal{N}{=}\,4$ supercharges generating the $\varepsilon$-transformations are computed to be
\begin{eqnarray}
Q^i &=& \mathrm{Tr}\Big( P\varphi^{i}-\frac{i}{3}\,\varphi_k[\varphi^{(i},\bar\varphi^{k)}]  -i\varphi_k\phi^{(iA}\phi{}^{k)}_A \Big)\,,
\nonumber \\
\bar Q_i &=& \mathrm{Tr}\Big( P\bar\varphi_{i} +
\frac{i}{3}\,\bar\varphi^k[\varphi_{(i},\bar\varphi_{k)}]  +i\bar\varphi^k\phi_{(i}^{A}\phi_{k)A} \Big)\,,
\label{Q-4}
\end{eqnarray}
where $P:=\{M,\dot M\}$ stands for the matrix momenta. Note that
$\mathrm{Tr}\left(\varphi_k[\varphi^{(i},\bar\varphi^{k)}]\right)=-\frac32\,\mathrm{Tr}\left(\bar \varphi^i\varphi^k \varphi_k\right)$,
$\mathrm{Tr}\left(\bar\varphi^k[\varphi_{(i},\bar\varphi_{k)}]\right)=-\frac32\,\mathrm{Tr}\left(\varphi_i\bar\varphi^k \bar\varphi_k\right)$.
The Noether charges associated with the odd $\eta$-transformations are
\begin{equation} \label{S-4}
S^i = \frac43\,\mathrm{Tr}\big( M^2\varphi^{i} \big) -tQ^i \,,
\qquad
\bar S_i = \frac43\,\mathrm{Tr}\big( M^2\bar\varphi_{i} \big) -t\bar Q_i\,.
\end{equation}

It is worth noting that the supercharges \p{Q-4} and \p{S-4} involve only physical fields and their momenta.
The corresponding transformations of auxiliary fields can be found using their algebraic equations of motion.

\subsection{Implicit ${\cal N}{=}\,4$, $d\,{=}\,1$ supersymmetry}

\quad\, As was shown in  \cite{F4scm}, in the one-particle case ($n\,{=}\,1$) the total action \p{act-total}
is invariant with respect to extra implicit  ${\cal N}{=}\,4$ supersymmetry transformations. The multiparticle (matrix)
generalization of these transformations \cite{ABC} reads
\be \label{trans-s8}
\delta_\xi \mathcal{M}=-\xi_{iA} \Psi^{iA}\,,\qquad  \delta_\xi \Psi^{iA}=\frac12\,\xi_k^A\left( D^i\bar D^k-\bar D^i D^k\right) \mathcal{M}\,,
\ee
where $\xi_{iA}$ are fermionic parameters. The superfield transformations \p{trans-s8} amount to the following ones for the component fields
\be \label{trans-c8}
\begin{array}{c}
\delta_\xi M=-\xi_{iA} \phi^{iA}\,,\qquad
\delta_\xi \varphi^{i}=\xi^{iA}F_A\,,\quad\delta_\xi \bar\varphi_{i}=-\xi_{iA}\bar F^A\,,\qquad
\delta_\xi A_{ik}=2\,\xi_{(iA} \dot\phi^{A}_{k)}\,, \\ [7pt]
\delta_\xi \phi^{iA}=i\xi^{iA}\dot M +i \xi^{A}_k A^{ik}\,,\qquad  \delta_\xi F^{A}=-2i\xi_k^A \dot\varphi^k\,,\quad
\delta_\xi \bar F_{A}= 2i\xi^k_A \dot{\bar\varphi}_k\,.
\end{array}
\ee
Thus the matrix ${\cal N}=4$ multiplets $({\bf 1, 4, 3})$ and $({\bf 0, 4, 4})$ in  the model under consideration
constitute together  $n^2$ matrix ${\cal N}{=}\,8$ multiplets $({\bf 1, 8, 7})$.

The variation of the total action \p{act-total} written in terms of components (that is, the sum of the component actions \p{4N-X-WZ}
and \p{coup-143-comp}) with respect to the transformations \p{trans-c8} is the integral
\be \label{var-c8-act}
\delta_\xi S_{(\mathcal{M}+\Psi)}=\int dt\, \dot\Lambda_\xi\,,
\ee
where
\begin{eqnarray}
\Lambda_\xi &=& \frac12\,\xi_{iA}\,\mathrm{Tr}\Big( -M\{\dot M,\phi^{iA}\} + M\{A^{ik},\phi^A_k \}+
iM\{\bar F^{A},\varphi^i \}- iM\{F^{A},\bar \varphi^i \}
\nonumber \\
&& \qquad\qquad\quad
+\ \frac{2i}{3}\,\phi^A_k\phi^{(iB}\phi^{k)}_B + 2i\phi^A_k[\varphi^{(i},\bar\varphi^{k)}] \Big)\,.
\label{Lambda-1}
\end{eqnarray}
The Noether charges of this hidden supersymmetry are then easily computed to be
\begin{equation}\label{Q-hidden}
\mathcal{Q}^{iA} = \mathrm{Tr}\Big( P\phi^{iA} +\frac{i}{3}\,\phi^A_k\phi^{(iB}\phi^{k)}_B
+i\,\phi^A_k[\varphi^{(i},\bar\varphi^{k)}] \Big)\,.
\end{equation}

In the next sections we will prove that the closure of the supersymmetry transformations \p{tr-comp-044}, \p{tr-comp-143}, \p{trans-c8}
generated by the supercharges \p{Q-4}, \p{S-4}, \p{Q-hidden} is just $\mathcal{N}{=}\,8$ conformal superalgebra $F(4)$
\cite{FRS,VP}.

\section{Harmonic variables in the Hermitian matrix model}

\quad\, The kinetic terms of bosonic and fermionic fields in the component actions \p{conf_v-13} and \p{Confpsi}
are not fully flat. In this section we extract, from the complete set of bosonic fields,  $n$ fields having flat kinetic terms.
The residual $n(n-1)$ bosonic variables are described by a non-trivial $d\,{=}\,1$ non-linear sigma models and admit
a suggestive interpretation as the target harmonics. After the appropriate redefinitions, the kinetic terms of all
fermionic fields will acquire the flat form.

The basic step in proving these assertions will be the spectral decomposition of the matrix ${M}$.
The Hermitian matrix ${M}={M}^\dagger$ is unitarily diagonalizable and
its eigen value-decomposition takes the form (see, for example, \cite{HornJons})
\begin{equation}\label{M-UY}
{M} = U \, Y \, U^\dagger \,,
\end{equation}
where $n\times n$ matrix $U$ is unitary,
\begin{equation}\label{Unit1}
U U^\dagger = I \,.
\end{equation}
In terms of the entries of $U$,
\begin{equation}\label{U-matrix}
U = \|u_{a}{}^{\beta}\| \,,\quad  U^\dagger = \|\bar u_{\alpha}{}^{b}\|\,,\qquad a=1,\ldots ,n\,,\quad  \alpha=1,\ldots ,n\,,
\end{equation}
$u_{a}^{\beta} \equiv u_{a}{}^{\beta}$, $\bar u_{\alpha}^{b} \equiv \bar u_{\alpha}{}^{b}$,
$\bar u_{\alpha}^{b}=(u_{b}^{\alpha})^*$, the unitarity condition amounts to the relations
\begin{equation}\label{u-unitary}
u_{a}^{\gamma}\bar u_{\gamma}^{b}=\delta_a^b\,,\qquad  u_{c}^{\alpha}\bar u_{\beta}^{c}=\delta_\beta^\alpha\,.
\end{equation}
The matrix $Y$ in the decomposition \p{M-UY} is a diagonal matrix,
\begin{equation}\label{Y-matrix}
Y = \|y_{\alpha}\delta_\alpha^{\beta}\|,
\end{equation}
with the real eigenvalues $y_{\alpha}=(y_{\alpha})^*$.
Thus, the components of the matrix ${M}$ defined in \p{M-UY} are expressed as
\begin{equation}\label{M-UY-comp}
{M}_a{}^b = \sum\limits_{\gamma=1}^N \, y_\gamma\, u_{a}^{\gamma}\,\bar u_{\gamma}^{b} \,.
\end{equation}
In this paper we will consider the option with unequal eigenvalues of the matrix $M$  \p{M-UY}, {\it i.e.} with
$y_\alpha\,{\neq}\, y_\beta$ for all $\alpha$, $\beta$.

Taking into account the diagonal form of the matrix $Y$ and the decomposition \p{M-UY}, we observe that
the components of the unitary matrix $U$ are defined up to local $[\mathrm{U}(1)]^n$ transformations
\begin{equation}\label{u-loc-trans}
u_{a}^{\beta} \ \to \ e^{i\vartheta_\beta} u_{a}^{\beta}\,,\qquad  \bar u_{\beta}^{a} \  \to  \  e^{-i\vartheta_\beta}\bar u_{\beta}^{b}\,,
\end{equation}
where $\vartheta_\alpha(t)$ ($\alpha \ {=}\, 1,\ldots , n$)  are local real parameters. Thus,
the matrix $U$ is defined up to the right local transformations
\begin{equation}\label{u-loc-matrix-trans}
U \ \to \ U h\,,
\end{equation}
where the matrix $h$ is diagonal with the components $e^{i\vartheta_\alpha}$, $\alpha \ {=}\, 1,\ldots , n$.
In a fixed gauge with respect to these right local shifts, the matrices $U$ involve $n^2 - n$ essential parameters
and so parametrize the cosets $\mathrm{U}(n)/H$ with the abelian stability subgroups
$H=\mathrm{U}_1(1)\otimes\ldots\otimes \mathrm{U}_n(1)$.
Then the variables $u_{a}^{\beta}$ and $\bar u_{\beta}^{a}$ can be interpreted as the
${\displaystyle \frac{\mathrm{U}(n)}{\mathrm{U}_1(1)\otimes\ldots\otimes \mathrm{U}_n(1)}}$ target harmonics.
Similar harmonics were considered in \cite{GIKOS-N3} for the case $n\,{=}\,3$ and in
\cite{Kallosh-1985,Bandos-1988} for arbitrary $n$. In the $n\,{=}\,2$ case we face just the target space analogs
of the standard ${\displaystyle \frac{\mathrm{SU}(2)}{\mathrm{U}(1)}}$
harmonics defined in \p{h-2-st} \cite{GIKOS,HSS}.

In accord with what has been said above, the $\mathrm{U}(n)$ transformations act on the indices $a$, $b$ whereas the indices $\alpha$, $\beta$
are subject to the $\mathrm{U}_1(1)\otimes\ldots\otimes \mathrm{U}_n(1)$ transformations. The harmonics $u_{a}^{\beta}$, $\bar u_{\beta}^{a}$
play the role of the bridges connecting the quantities with different types of symmetry, $\mathrm{U}(n)$ and  $[\mathrm{U}(1)]^n$.

Let us rewrite the total action \p{act-total} (the sum of the component actions \p{4N-X-WZ} and \p{coup-143-comp})
in terms of the variables $y_\alpha$, $u_{a}^{\beta}$, $\bar u_{\beta}^{a}$. The crucial role will be played by the relation~\cite{Poly-rev}
\begin{equation}\label{dot-M}
\dot M = U \left(\dot Y + [\mathcal{K},Y] \right) U^\dagger \,,
\end{equation}
where
\begin{equation}\label{math-K}
\mathcal{K}:= \frac12\left(U^\dagger \dot U - \dot U^\dagger U \right) ,\qquad
\mathcal{K}_\alpha{}^\beta= \frac12\left(\bar u^c_\alpha \dot u_c^\beta - \dot{\bar u}^c_\alpha u_c^\beta \right) .
\end{equation}

The first term in the Lagrangian of the action \p{4N-X-WZ} takes the form
\begin{eqnarray}
\mathrm{Tr}\Big( M\dot M\dot M\Big) &=& \mathrm{Tr}\Big( Y\dot Y\dot Y\Big)+\mathrm{Tr}\Big( Y[\mathcal{K},Y]^2\Big)
\nonumber \\
&=&
\sum\limits_{\alpha} y_\alpha \dot y_\alpha \dot y_\alpha -
\frac12 \sum\limits_{\alpha\neq\beta}(y_\alpha+y_\beta)(y_\alpha-y_\beta)^2\mathcal{K}_\alpha{}^\beta\mathcal{K}_\beta{}^\alpha\,.
\label{4N-X-1-comp}
\end{eqnarray}
The remaining terms in the actions \p{4N-X-WZ} and \p{coup-143-comp} can be simplified after
introducing new fermionic matrix variables,
\begin{equation}\label{new-ferni-matr}
\underset{{}^{{}^\sim}}{\phi}{}^A_k := U^\dagger\, {\phi}{}^A_k\, U\,,\qquad
\underset{{}^{{}^\sim}}{\varphi}{}^k := U^\dagger\, {\varphi}{}^k\, U\,,\qquad
\underset{{}^{{}^\sim}}{\bar\varphi}{}_k := U^\dagger\, {\bar\varphi}{}_k\, U\,,
\end{equation}
and new bosonic matrix variables,
\begin{equation}\label{new-bose-matr}
\underset{{}^{{}^\sim}}{A}{}^{ik} := U^\dagger\, A^{ik}\, U\,,\qquad
\underset{{}^{{}^\sim}}{F}{}^A := U^\dagger\, {F}{}^A\, U\,,\qquad
\underset{{}^{{}^\sim}}{\bar F}{}_A := U^\dagger\, {\bar F}{}_A\, U\,.
\end{equation}
New matrix quantities \p{new-ferni-matr}, \p{new-bose-matr} carry the $\mathrm{U}_1(1)\otimes\ldots\otimes \mathrm{U}_n(1)$ indices:
$(\underset{{}^{{}^\sim}}{A}{}^{ik})_\alpha{}^\beta$,
$(\underset{{}^{{}^\sim}}{\phi}{}^A_k)_\alpha{}^\beta$, $(\underset{{}^{{}^\sim}}{\varphi}{}^k)_\alpha{}^\beta$, \textit{etc}.

The definitions \p{new-ferni-matr}, \p{new-bose-matr}  imply the relations
\begin{equation}\label{new-ferni-matr-der}
U^\dagger\, \dot{\phi}{}^A_k\, U=\underset{{}^{{}^\sim}}{\dot\phi}{}^A_k +[\mathcal{K},\underset{{}^{{}^\sim}}{\phi}{}^A_k] \,,\qquad
U^\dagger\, \dot{\varphi}{}^k\, U=\underset{{}^{{}^\sim}}{\dot\varphi}{}^k +[\mathcal{K},\underset{{}^{{}^\sim}}{\varphi}{}^k] \,,\qquad
U^\dagger\, \dot{\bar\varphi}{}_k\, U=\underset{{}^{{}^\sim}}{\dot{\bar\varphi}}{}_k +[\mathcal{K},\underset{{}^{{}^\sim}}{\bar\varphi}{}_k] \,.
\end{equation}
Using them, the total action  \p{act-total} can be rewritten in terms of the new variables \p{Y-matrix}, \p{U-matrix}, \p{new-ferni-matr},
\p{new-bose-matr} as
\bea \label{total-comp-new}
\!\!S_{(\mathcal{M}+\Psi)} \!\!
&\!\! = \!\!&\!\!  \displaystyle{\int} dt\,
\mathrm{Tr}\left( Y\dot Y\dot Y +  Y[\mathcal{K},Y]^2
+ \frac{1}{2}\, Y \underset{{}^{{}^\sim}}{A}{}^{ik} \underset{{}^{{}^\sim}}{A}{}_{ik}\,
+  \frac{1}{2} \,Y\{\underset{{}^{{}^\sim}}{F}{}^A,\underset{{}^{{}^\sim}}{\bar F}{}_A\} \right) \nonumber \\
\!\!&&\!\!  -\frac{i}{2}\displaystyle{\int} dt\,
\mathrm{Tr}\left(\{Y, \underset{{}^{{}^\sim}}{\bar\varphi}{}^k\} \underset{{}^{{}^\sim}}{\dot\varphi}{}_k
+\{Y,\underset{{}^{{}^\sim}}{\varphi}{}_k\} \underset{{}^{{}^\sim}}{\dot{\bar\varphi}}{}^k
+ \{Y,\underset{{}^{{}^\sim}}{\phi}{}^{iA}\}\underset{{}^{{}^\sim}}{\dot\phi}{}_{iA}
\right) \nonumber \\
\!\!&&\!\!  -\frac{i}{2}\displaystyle{\int} dt\,
\mathrm{Tr}\left(\{Y, \underset{{}^{{}^\sim}}{\bar\varphi}{}^k\} [\mathcal{K},\underset{{}^{{}^\sim}}{\varphi}{}_k]
+\{Y,\underset{{}^{{}^\sim}}{\varphi}{}_k\} [\mathcal{K},\underset{{}^{{}^\sim}}{\bar\varphi}{}^k]
+ \{Y,\underset{{}^{{}^\sim}}{\phi}{}^{iA}\}[\mathcal{K},\underset{{}^{{}^\sim}}{\phi}{}_{iA}]
\right) \nonumber \\
\!\!&&\!\! +\frac12 \int dt\, \mathrm{Tr}\left(
i \Big([\underset{{}^{{}^\sim}}{\varphi}{}_{i},\underset{{}^{{}^\sim}}{\bar\varphi}{}_{k}]
+ \underset{{}^{{}^\sim}}{\phi}{}^A_i\underset{{}^{{}^\sim}}{\phi}{}_{kA} \Big)\underset{{}^{{}^\sim}}{A}{}^{ik}
+[\underset{{}^{{}^\sim}}{\bar\varphi}{}^k, \underset{{}^{{}^\sim}}{\phi}{}_{kA}] \underset{{}^{{}^\sim}}{F}{}^A
- [\underset{{}^{{}^\sim}}{\varphi}{}_k ,\underset{{}^{{}^\sim}}{\phi}{}^{kA}] \underset{{}^{{}^\sim}}{\bar F}{}_A \right) .
\eea
The fields $\underset{{}^{{}^\sim}}{A}{}^{ik}$, $\underset{{}^{{}^\sim}}{F}{}^A$, $\underset{{}^{{}^\sim}}{\bar F}{}_A$
are auxiliary. In the next section, before performing the Hamiltonian analysis, we will eliminate them
and diagonalize the kinetic terms for the bosonic $y$-variables and for the fermionic ones.

\section{The physical-variable form of the system}

\subsection{On-shell action}

\quad\, Elimination of the auxiliary fields $\underset{{}^{{}^\sim}}{A}{}^{ik}$
and $\underset{{}^{{}^\sim}}{F}{}^A$, $\underset{{}^{{}^\sim}}{\bar F}{}_A$ in the component action \p{total-comp-new}
by their equations of motion,
\be
\begin{array}{c}
\underset{{}^{{}^\sim}}{A}{}_{ik}{}_\alpha{}^\beta= -i (y_\alpha+y_\beta)^{-1}
\big([\underset{{}^{{}^\sim}}{\varphi}{}_{(i},\underset{{}^{{}^\sim}}{\bar\varphi}{}_{k)}]
+ \underset{{}^{{}^\sim}}{\phi}{}^A_{(i}\underset{{}^{{}^\sim}}{\phi}{}_{k)A}\big){}_\alpha{}^\beta\,,
\\ [7pt]
\underset{{}^{{}^\sim}}{F}{}^A{}_\alpha{}^\beta =
(y_\alpha+y_\beta)^{-1}[\underset{{}^{{}^\sim}}{\varphi}{}_k ,\underset{{}^{{}^\sim}}{\phi}{}^{kA}]{}_\alpha{}^\beta\,,\qquad
\underset{{}^{{}^\sim}}{\bar F}{}_A{}_\alpha{}^\beta =-(y_\alpha+y_\beta)^{-1}
[\underset{{}^{{}^\sim}}{\bar\varphi}{}^k, \underset{{}^{{}^\sim}}{\phi}{}_{kA}]{}_\alpha{}^\beta
\end{array},
\ee
produces the 4-fermionic terms.
As the result, we obtain the total on-shell superconformal action in the form
\bea
\!\!S_{(\mathcal{M}+\Psi)} \!\!
&\!\! = \!\!&\!\!  \displaystyle{\int} dt
\left( \sum\limits_{\alpha} y_\alpha \dot y_\alpha \dot y_\alpha -
\frac12 \sum\limits_{\alpha,\beta}(y_\alpha+y_\beta)(y_\alpha-y_\beta)^2\mathcal{K}{}_\alpha{}^\beta\mathcal{K}{}_\beta{}^\alpha
\right) \nonumber \\
\!\!&&\!\!  -\frac{i}{2}\displaystyle{\int} dt\,
\sum\limits_{\alpha,\beta}(y_\alpha+y_\beta)
\left(\underset{{}^{{}^\sim}}{\bar\varphi}{}^k{}_\alpha{}^\beta \underset{{}^{{}^\sim}}{\dot\varphi}{}_k{}_\beta{}^\alpha
+ \underset{{}^{{}^\sim}}{\varphi}{}_k{}_\alpha{}^\beta \underset{{}^{{}^\sim}}{\dot{\bar\varphi}}{}^k{}_\beta{}^\alpha
+  \underset{{}^{{}^\sim}}{\phi}{}^{iA}{}_\alpha{}^\beta \underset{{}^{{}^\sim}}{\dot\phi}{}_{iA}{}_\beta{}^\alpha
\right) \nonumber \\
\!\!&&\!\!  + \frac{i}{2}\displaystyle{\int} dt\,
\sum\limits_{\alpha,\beta,\gamma}
(y_\alpha+y_\beta)
\left( \underset{{}^{{}^\sim}}{\varphi}{}_k{}_\alpha{}^\beta   \underset{{}^{{}^\sim}}{\bar\varphi}{}^k{}_\beta{}^\gamma
+  \underset{{}^{{}^\sim}}{\bar\varphi}{}^k{}_\alpha{}^\beta   \underset{{}^{{}^\sim}}{\varphi}{}_k{}_\beta{}^\gamma
+ \underset{{}^{{}^\sim}}{\phi}{}^{iA}{}_\alpha{}^\beta \underset{{}^{{}^\sim}}{\phi}{}_{iA}{}_\beta{}^\gamma
\right) \mathcal{K}{}_\gamma{}^\alpha\nonumber \\
\!\!&&\!\!  - \frac{i}{2}\displaystyle{\int} dt\,
\sum\limits_{\alpha,\beta,\gamma}
(y_\alpha+y_\beta)
\left( \underset{{}^{{}^\sim}}{\varphi}{}_k{}_\alpha{}^\beta   \underset{{}^{{}^\sim}}{\bar\varphi}{}^k{}_\gamma{}^\alpha
+  \underset{{}^{{}^\sim}}{\bar\varphi}{}^k{}_\alpha{}^\beta   \underset{{}^{{}^\sim}}{\varphi}{}_k{}_\gamma{}^\alpha
+ \underset{{}^{{}^\sim}}{\phi}{}^{iA}{}_\alpha{}^\beta \underset{{}^{{}^\sim}}{\phi}{}_{iA}{}_\gamma{}^\alpha
\right) \mathcal{K}{}_\beta{}^\gamma\nonumber \\
\!\!&&\!\! +\frac12 \int dt\,
\sum\limits_{\alpha,\beta}
(y_\alpha+y_\beta)^{-1} [\underset{{}^{{}^\sim}}{\varphi}{}_k ,\underset{{}^{{}^\sim}}{\phi}{}^{kA}]{}_\alpha{}^\beta
[\underset{{}^{{}^\sim}}{\bar\varphi}{}^k, \underset{{}^{{}^\sim}}{\phi}{}_{kA}]{}_\beta{}^\alpha  \nonumber \\
\!\!&&\!\!  + \frac{1}{4}\displaystyle{\int} dt\,
\sum\limits_{\alpha,\beta} (y_\alpha+y_\beta)^{-1}
\big([\underset{{}^{{}^\sim}}{\varphi}{}^{i},\underset{{}^{{}^\sim}}{\bar\varphi}{}^{k}]
+ \underset{{}^{{}^\sim}}{\phi}{}^{iA}\underset{{}^{{}^\sim}}{\phi}{}_{A}^k\big){}_\alpha{}^\beta
\big([\underset{{}^{{}^\sim}}{\varphi}{}_{(i},\underset{{}^{{}^\sim}}{\bar\varphi}{}_{k)}]
+ \underset{{}^{{}^\sim}}{\phi}{}^B_{(i}\underset{{}^{{}^\sim}}{\phi}{}_{k)B}\big){}_\beta{}^\alpha .
\label{total-comp-new-expl-on}
\eea

Redefining field variables as\footnote{Similar fractional-degree redefinitions of the target coordinates appeared in \cite{FI-2015,F4-l}.}
\begin{equation}\label{4N-nZ}
x_\alpha = \frac{2\sqrt{2}}{3}\, (y_\alpha)^{3/2} \,, \qquad \psi_{k}{}_\alpha{}^\beta  =
(y_\alpha+y_\beta)^{1/2}\, \underset{{}^{{}^\sim}}{\varphi}{}_k{}_\alpha{}^\beta \,, \qquad
\chi^{iA}{}_\alpha{}^\beta  = (y_\alpha+y_\beta)^{1/2}\,\underset{{}^{{}^\sim}}{\phi}{}^{iA}{}_\alpha{}^\beta \,,
\end{equation}
we cast the action \p{total-comp-new-expl-on} in the more convenient form
\bea
\label{total-comp-new-expl-on-red}
S_{(\mathcal{M}+\Psi)} &= & \int dt \ L_{(\mathcal{M}+\Psi)} \,,\\ [6pt]
\nonumber
L_{(\mathcal{M}+\Psi)}
& = &  \frac12\,
\sum\limits_{\alpha} \dot x_\alpha \dot x_\alpha \,-\,\frac{i}{2}\,
\sum\limits_{\alpha,\beta}
\left(\bar\psi^k{}_\alpha{}^\beta \dot\psi_k{}_\beta{}^\alpha
+ \psi{}_k{}_\alpha{}^\beta \dot{\bar\psi}{}^k{}_\beta{}^\alpha
+  \chi{}^{iA}{}_\alpha{}^\beta \dot\chi{}_{iA}{}_\beta{}^\alpha
\right) \\ [5pt]
&&  -\,
\frac{9}{16}\, \sum\limits_{\alpha,\beta}\Delta^{+}_{\alpha\beta}\,(\Delta^{-}_{\alpha\beta})^2\,
\mathcal{K}{}_\alpha{}^\beta\mathcal{K}{}_\beta{}^\alpha
+ \,\frac{i}{2}\,
\sum\limits_{\alpha,\beta}\Omega_{\alpha}{}^{\beta} \mathcal{K}{}_\beta{}^\alpha  \,+ \,L_{(4-f)}\,.
\label{total-comp-new-expl-on-red-L}
\eea
Here
\begin{equation}\label{Dela-def}
\Delta^{\pm}_{\alpha\beta}:=x_\alpha^{2/3}\pm x_\beta^{2/3}\,,
\end{equation}
\begin{equation}\label{Om-def}
\Omega_{\alpha}{}^{\beta}:=
\sum\limits_{\gamma}
\frac{\Delta^{+}_{\alpha\gamma}+\Delta^{+}_{\beta\gamma}}{\left(\Delta^{+}_{\alpha\gamma}\Delta^{+}_{\beta\gamma}\right)^{1/2}}\,
\Big( \psi{}_k{}_\alpha{}^\gamma  \bar\psi{}^k{}_\gamma{}^\beta
+  \bar\psi{}^k{}_\alpha{}^\gamma  \psi{}_k{}_\gamma{}^\beta
+ \chi{}^{iA}{}_\alpha{}^\gamma \chi{}_{iA}{}_\gamma{}^\beta
\Big)
\end{equation}
and the 4-fermionic term $L_{(4-f)}$ reads
\bea
\nonumber
L_{(4-f)}
& = &  \frac{2}{9}
\sum\limits_{\alpha,\beta,\gamma,\delta} \frac
{1}
{\Delta^{+}_{\alpha\beta} (\Delta^{+}_{\alpha\gamma}){}^{1/2}(\Delta^{+}_{\alpha\delta}){}^{1/2} (\Delta^{+}_{\beta\gamma}){}^{1/2} (\Delta^{+}_{\beta\delta}){}^{1/2}}\,
\Bigg\{ \chi{}^{iA}{}_\alpha{}^\gamma \chi{}_{A}^k{}_\gamma{}^\beta
\chi{}^B_{(i}{}_\beta{}^\delta \chi{}_{k)B}{}_\delta{}^\alpha
\\
&&  \ \ \ \
+\, \big(\psi{}^{i}{}_\alpha{}^\gamma\bar\psi{}^{k}{}_\gamma{}^\beta - \bar\psi{}^{k}{}_\alpha{}^\gamma\psi{}^{i}{}_\gamma{}^\beta \big)
\big(\psi{}_{(i}{}_\beta{}^\delta \bar\psi{}_{k)}{}_\delta{}^\alpha - \bar\psi{}_{(i)}{}_\beta{}^\delta \psi{}_{k)}{}_\delta{}^\alpha \big)
\nonumber\\ [7pt]
&&  \ \ \ \
+\, 2\big(\psi{}^{i}{}_\alpha{}^\gamma\bar\psi{}^{k}_\gamma{}^\beta- \bar\psi{}^{k}{}_\alpha{}^\gamma\psi{}^{i}_\gamma{}^\beta\big)
\chi{}^A_{(i}{}_\beta{}^\delta  \chi{}_{k)A}{}_\delta{}^\alpha
\nonumber \\
&&  \ \ \ \
+\, 2\big( \psi{}_k{}_\alpha{}^\gamma \chi{}^{kA}{}_\gamma{}^\beta - \chi{}^{kA}{}_\alpha{}^\gamma \psi{}_k{}_\gamma{}^\beta \big)
\big( \bar\psi{}^l{}_\beta{}^\delta \chi{}_{lA}{}_\delta{}^\alpha - \chi{}_{lA}{}_\beta{}^\delta \bar\psi{}^l{}_\delta{}^\alpha  \big)
\Bigg\}\,.
\label{4-ferm-comp-new-expl-on-red}
\eea
The Lagrangian \p{total-comp-new-expl-on-red-L} contains flat kinetic terms for $x$-variables and fermions.
Harmonics are dynamical in this model: their second order kinetic term is proportional to $\mathcal{K}^2$ which
is just the relevant target space sigma-model metric.

In the one-particle  case ($n\,{=}\,1$), the action \p{total-comp-new-expl-on-red} is reduced to the on-shell
action from ref.~\cite{F4scm} and, at $\alpha\,{=}\,{-}1/3$, to the action from ref.~\cite{FI-2015,Gal-2017}.\footnote{
Such type of interaction of one multiplet $({\bf 1, 4, 3})$ with one multiplet $({\bf 0, 4, 4})$ for the special case of
$\mathrm{SU}(1,1|2)$ superconformal symmetry was considered in \cite{Gal-2015}.}

\subsection{Hamiltonian and harmonic constraints}

\quad\, The Lagrangian \p{total-comp-new-expl-on-red-L} yields the following explicit expressions for the momenta:
\be\label{mom-x}
p_\alpha=\frac{\partial L_{(\mathcal{M}+\Psi)}}{\partial \dot x_\alpha} \ = \ \dot x_\alpha\,,
\ee
\be
\label{mom-gr}
\begin{array}{rcl}
{\displaystyle \Pi_k{}_\alpha{}^\beta=\frac{\partial^{\,l} L_{(\mathcal{M}+\Psi)}}{\partial \dot \psi^k{}_\beta{}^\alpha}}
&=&{\displaystyle \frac{i}{2}\,\bar\psi_k{}_\alpha{}^\beta\,,\qquad\quad
\bar \Pi^k{}_\alpha{}^\beta=\frac{\partial^{\,l} L_{(\mathcal{M}+\Psi)}}{\partial \dot {\bar\psi}_k{}_\beta{}^\alpha}
\ = \ \frac{i}{2}\,\psi{}^k{}_\alpha{}^\beta\,,}\\[7pt]
{\displaystyle \Pi_{kA}{}_\alpha{}^\beta=\frac{\partial^{\,l} L_{(\mathcal{M}+\Psi)}}{\partial \dot {\chi}^{kA}{}_\beta{}^\alpha}}
&=&{\displaystyle -\frac{i}{2}\,\chi{}_{kA}{}_\alpha{}^\beta \,, }
\end{array}
\ee
\be
\begin{array}{rcl}
{\displaystyle \pi_\alpha^b=\frac{\partial L_{(\mathcal{M}+\Psi)}}{\partial \dot u_b^\alpha}}&=&
{\displaystyle-\frac{1}{16}\, \sum\limits_{\beta} \Big(9\,\Delta^{+}_{\alpha\beta}\,(\Delta^{-}_{\alpha\beta})^2\,
\mathcal{K}{}_\alpha{}^\beta \, - \,4i\,\Omega_{\alpha}{}^{\beta}\Big)  \bar u_\beta^b\,,}\\ [7pt]
{\displaystyle \bar\pi^\alpha_b=\frac{\partial L_{(\mathcal{M}+\Psi)}}{\partial \dot {\bar u}_\alpha^b}} &=&
{\displaystyle\frac{1}{16}\, \sum\limits_{\beta} \Big(9\,\Delta^{+}_{\alpha\beta}\,(\Delta^{-}_{\alpha\beta})^2\,
\mathcal{K}{}_\beta{}^\alpha \, - \,4i\,\Omega_{\beta}{}^{\alpha}\Big) u^\beta_b\,. }
\end{array}
\label{mom-harm}
\ee
The canonical Hamiltonian
\begin{equation}\label{Ham-can}
H=p_\alpha\dot x_\alpha+\Pi_k{}_\alpha{}^\beta \dot \psi^k{}_\beta{}^\alpha+\bar \Pi^k{}_\alpha{}^\beta\dot {\bar\psi}_k{}_\beta{}^\alpha
+\Pi_{kA}{}_\alpha{}^\beta \dot {\chi}^{kA}{}_\beta{}^\alpha
+\pi_\alpha^b \dot u_b^\alpha + \bar \pi^\alpha_b \dot {\bar u}_\alpha^b -L_{(\mathcal{M}+\Psi)}
\end{equation}
takes the form
\begin{equation}\label{Hamil}
H=\frac12\,\sum\limits_{\alpha} p_\alpha p_\alpha \,-\,
\frac{4}{9}\, \sum\limits_{\alpha\neq\beta}
\frac{\mathcal{D}_\alpha^\beta \mathcal{D}_\beta^\alpha}{\Delta^{+}_{\alpha\beta}\,(\Delta^{-}_{\alpha\beta})^2}\,
\,-\, \,L_{(4-f)}\,,
\end{equation}
where
\begin{equation}\label{D2-def}
\mathcal{D}^\alpha_\beta:=D^\alpha_\beta-
\frac{i}{2}\,\Omega_\beta{}^\alpha\,,\qquad D^\alpha_\beta:=u^\alpha_c\pi_\beta^c-\bar u_\beta^c\bar\pi^\alpha_c
\end{equation}
and $\Omega_\beta{}^\alpha$ was defined in \p{Om-def}.
We point out that only the components $\mathcal{K}{}_\alpha{}^\beta$ at $\alpha\neq\beta$ were used in the calculation
of the Hamiltonian \p{Hamil}. These quantities are expressed as
\begin{equation}\label{K-eq}
\mathcal{K}{}_\alpha{}^\beta= -\frac{8\,\mathcal{D}{}_\alpha{}^\beta}{9 \Delta^{+}_{\alpha\beta}\,(\Delta^{-}_{\alpha\beta})^2}  \,.
\end{equation}
Remind that we consider the case with $y_\alpha\,{\neq}\, y_\beta$ for any $\alpha\,{\neq}\,\beta$ and so
all $\Delta^{-}_{\alpha\beta}$ in \p{K-eq} are non-vanishing.

The expressions \p{mom-gr} produce the standard second class constraints for odd variables.
After introducing Dirac brackets for them, odd momenta $\Pi_k{}_\alpha{}^\beta$, $\bar \Pi^k{}_\alpha{}^\beta$, $\Pi_{kA}{}_\alpha{}^\beta$
are removed from the phase space.
The non-vanishing canonical Dirac brackets for the residual variables (at equal times) are
\begin{equation}\label{CDB1}
\{x_\alpha, p_\beta\}^*= \delta_{\alpha\beta}\,, \qquad
\{u^\alpha_a, \pi_\beta^b\}^*= \delta^b_a \delta^\alpha_\beta\,,\qquad
\{\bar u_\alpha^a, \bar\pi^\beta_b\}^*= \delta_b^a \delta_\alpha^\beta\,,
\end{equation}
\begin{equation}\label{CDB2}
\{\psi{}^i{}_\alpha{}^\beta, \bar\psi_k{}_\gamma{}^\delta\}^*= -i\,\delta^i_k \delta^\delta_\alpha \delta^\beta_\gamma\,,\qquad
\{\chi^{iA}{}_\alpha{}^\beta, \chi^{jB}{}_\gamma{}^\delta\}^*= i\,\epsilon^{ij}\epsilon^{AB} \delta^\delta_\alpha \delta^\beta_\gamma\,,
\end{equation}
where $\epsilon_{12} = \epsilon^{21} = 1$.

Taking into account \p{u-unitary}, we note that in the expressions for the harmonic momenta \p{mom-harm} the following conditions
are implicitly used
\begin{equation}\label{cond-1}
u^\alpha_c\pi_\beta^c=-\bar u_\beta^c\bar\pi^\alpha_c \qquad \mbox{($\alpha, \beta$ are arbitrary; sum over $c$)}\,,
\end{equation}
\begin{equation}\label{cond-2}
u^\alpha_c\pi_\alpha^c-\bar u_\alpha^c\bar\pi^\alpha_c=
i\sum\limits_{\gamma}
\Big( \psi{}_k{}_\alpha{}^\gamma  \bar\psi{}^k{}_\gamma{}^\alpha
+  \bar\psi{}^k{}_\alpha{}^\gamma  \psi{}_k{}_\gamma{}^\alpha
+ \chi{}^{iA}{}_\alpha{}^\gamma \chi{}_{iA}{}_\gamma{}^\alpha
\Big)
 \quad \mbox{(no sum over $\alpha$; sum over $c$)}\,.
\end{equation}
As a consequence, the considered system possesses $n$ harmonic constraints
\begin{equation}\label{constr-1}
\mathcal{D}_\alpha:=u^\alpha_c\pi_\alpha^c-\bar u_\alpha^c\bar\pi^\alpha_c-
i\sum\limits_{\gamma}
\Big( \psi{}_k{}_\alpha{}^\gamma  \bar\psi{}^k{}_\gamma{}^\alpha
+  \bar\psi{}^k{}_\alpha{}^\gamma  \psi{}_k{}_\gamma{}^\alpha
+ \chi{}^{iA}{}_\alpha{}^\gamma \chi{}_{iA}{}_\gamma{}^\alpha
\Big)\approx 0
\end{equation}
(no sum over $\alpha$; sum over $c$), $n^2$ harmonic constraints
\begin{equation}\label{constr-2b}
G^\alpha_\beta:=u^\alpha_c\pi_\beta^c+\bar u_\beta^c\bar\pi^\alpha_c\approx 0\,,
\end{equation}
($\alpha$ and $\beta$ are arbitrary, summation over $c$) and $n^2$  kinematic constraints \p{u-unitary}
\begin{equation}\label{constr-2c}
g^\alpha_\beta:=u_{c}^{\alpha}\bar u_{\beta}^{c}-\delta_\beta^\alpha\approx 0\,.
\end{equation}
It should be pointed out that the quantities $\mathcal{D}_\alpha$ appearing in \p{constr-1} coincide with $\mathcal{D}^\alpha_\beta$ from \p{D2-def} at $\alpha{=}\beta$:
$D_\alpha{=}D_\alpha^\alpha$.

The quantities $D^\alpha_\beta$ defined in \p{D2-def} form $u(n)$ algebra with respect to Dirac brackets~\p{CDB2},
\begin{equation}\label{D-DB}
\{D^\alpha_\beta,D^\gamma_\delta\}^*= \delta^{\alpha}_{\delta}D^\gamma_\beta- \delta^\gamma_\beta D^{\alpha}_{\delta}\,,
\end{equation}
and commute, in a weak sense, with the quantities $G^\alpha_\beta$, defined in \p{constr-2b},
\begin{equation}\label{DG-DB}
\{G^\alpha_\beta,D^\gamma_\delta\}^*= \delta^{\alpha}_{\delta}G^\gamma_\beta- \delta^\gamma_\beta G^{\alpha}_{\delta}\,.
\end{equation}

The non-vanishing Dirac brackets of the constraints  \p{constr-2b} and \p{constr-2c} are
\begin{equation}\label{GG-DB}
\{G^\alpha_\beta,G^\gamma_\delta\}^*= \delta^{\alpha}_{\delta}D^\gamma_\beta- \delta^\gamma_\beta D^{\alpha}_{\delta}\,,
\end{equation}
\begin{equation}\label{gG-DB}
\{g^\alpha_\beta,G^\gamma_\delta\}^*= 2\delta^{\alpha}_{\delta}\delta^\gamma_\beta +\delta^{\alpha}_{\delta}g^\gamma_\beta
+ \delta^\gamma_\beta g^{\alpha}_{\delta}\,.
\end{equation}
Therefore, $n$ constraints $\mathcal{D}_\alpha$ \p{constr-1} are first class, whereas
the constraints $G^\alpha_\beta$ \p{constr-2b} and $g^\alpha_\beta$ \p{constr-2c} form $n^2$ pairs
of second class constraints.

We take account of the second class constraints \p{constr-2b}, \p{constr-2c}
by introducing Dirac brackets for them:
\bea
\nonumber
\{A,B\}^{**}
& = &  \{A,B\}^{*}+\frac12\,\{A,g^\alpha_\beta\}^{*}\{G_\alpha^\beta,B\}^{*} -\frac12\,\{A,G^\alpha_\beta\}^{*}\{g_\alpha^\beta,B\}^{*}\\ [5pt]
&&  -\,\frac14\,\{A,g^\alpha_\beta\}^{*}D_\alpha^\gamma\{g_\gamma^\beta,B\}^{*}
+\frac14\,\{A,g^\alpha_\beta\}^{*}D_\gamma^\beta\{g_\alpha^\gamma,B\}^{*}\,.
\label{DB-2}
\eea
The bracket  \p{DB-2} gives the necessary conditions $\{A,g^\alpha_\beta\}^{**}=\{A,G^\alpha_\beta\}^{**}=0$
for an arbitrary phase variable $A$.
For the non-harmonic variables new Dirac brackets coincide with the brackets  \p{CDB1},  \p{CDB2}:
\begin{equation}\label{CDB-2-nonh}
\{x_\alpha, p_\beta\}^{**}= \delta_{\alpha\beta}\,, \quad
\{\psi{}^i{}_\alpha{}^\beta, \bar\psi_k{}_\gamma{}^\delta\}^{**}= -i\,\delta^i_k \delta^\delta_\alpha \delta^\beta_\gamma\,,\quad
\{\chi^{iA}{}_\alpha{}^\beta, \chi^{jB}{}_\gamma{}^\delta\}^{**}= i\,\epsilon^{ij}\epsilon^{AB} \delta^\delta_\alpha \delta^\beta_\gamma\,.
\end{equation}
The new Dirac brackets for the $u(n)$ generators \p{D2-def} also retain the old form:
\begin{equation}\label{D-DB-2}
\{D^\alpha_\beta,D^\gamma_\delta\}^{**}= \delta^{\alpha}_{\delta}D^\gamma_\beta- \delta^\gamma_\beta D^{\alpha}_{\delta}\,.
\end{equation}
In what follows, this property will be crucial  for analyzing the superconformal symmetry.

\section{$\mathcal{N}{=}\,8$ supersymmetry generators}

\subsection{Odd generators of the  superalgebra $D(2,1;\alpha{=}{-}1/3)$}

\quad\, Using the relations \p{M-UY}, \p{dot-M}, \p{new-ferni-matr} and \p{4N-nZ}
we find that the generators of the
$\mathcal{N}{=}\,4$ supersymmetry \p{Q-4} take  the following form in the new variables
\begin{eqnarray}
Q^i &=& \sum\limits_{\alpha} p_\alpha \psi^i{}_{\alpha}{}^{\alpha}
+\frac{2\sqrt{2}}{3}\,\sum\limits_{\alpha\neq\beta}
\frac{\mathcal{D}_\alpha^\beta \psi{}^i{}_\beta{}^\alpha}
{(\Delta^{+}_{\alpha\beta})^{1/2}\Delta^{-}_{\alpha\beta} }
\nonumber \\
&& -\frac{i2\sqrt{2}}{9}\,
\sum\limits_{\alpha,\beta,\gamma}
\frac{\psi_k{}_\alpha{}^\beta \Big(\psi^{(i}{}_\beta{}^\gamma\bar\psi^{k)}{}_\gamma{}^\alpha
-\bar\psi^{(i}{}_\beta{}^\gamma\psi^{k)}{}_\gamma{}^\alpha+3\,\chi^{(iA}{}_\beta{}^\gamma\chi{}^{k)}_A{}_\gamma{}^\alpha \Big)}
{(\Delta^{+}_{\alpha\beta})^{1/2} (\Delta{}_{\alpha\gamma}^{+})^{1/2}(\Delta^{+}_{\beta\gamma})^{1/2}}\,,
\nonumber \\
\bar Q_i &=& \sum\limits_{\alpha} p_\alpha \bar\psi_i{}_{\alpha}{}^{\alpha}
+\frac{2\sqrt{2}}{3}\,\sum\limits_{\alpha\neq\beta}
\frac{\mathcal{D}_\alpha^\beta \bar\psi{}_i{}_\beta{}^\alpha}
{(\Delta^{+}_{\alpha\beta})^{1/2}\Delta^{-}_{\alpha\beta} }
\nonumber \\
&& +\frac{i2\sqrt{2}}{9}\,
\sum\limits_{\alpha,\beta,\gamma}
\frac{\bar\psi^k{}_\alpha{}^\beta \Big(\psi_{(i}{}_\beta{}^\gamma\bar\psi_{k)}{}_\gamma{}^\alpha
-\bar\psi_{(i}{}_\beta{}^\gamma\psi_{k)}{}_\gamma{}^\alpha+3\,\chi^A_{(i}{}_\beta{}^\gamma\chi{}_{k)A}{}_\gamma{}^\alpha \Big)}
{(\Delta^{+}_{\alpha\beta})^{1/2} (\Delta{}_{\alpha\gamma}^{+})^{1/2}(\Delta^{+}_{\beta\gamma})^{1/2}}\,,
\label{Q-4-n}
\end{eqnarray}
where $p_\alpha= \dot x_\alpha$ as in \p{mom-x} and $\Delta^{\pm}_{\alpha\beta}$ were defined in \p{Dela-def}.
The generators of the superconformal boosts \p{S-4} are
\begin{equation}\label{S-4-n}
{S}^i =\sum\limits_{\alpha} x_\alpha \psi^i{}_{\alpha}{}^{\alpha} - t\,{Q}^i,\qquad
\bar{S}_i=\sum\limits_{\alpha} x_\alpha \bar\psi_i{}_{\alpha}{}^{\alpha}-t\,\bar{Q}_i\,.
\end{equation}

With taking into account the relations \p{M-UY}, \p{dot-M}, \p{new-ferni-matr} and \p{4N-nZ},
the generators of the second $\mathcal{N}{=}\,4$ supersymmetry \p{Q-hidden} acquire  the following form in the new variables
\begin{eqnarray}
\mathcal{Q}^{iA} &=& \sum\limits_{\alpha} p_\alpha \chi^{iA}{}_{\alpha}{}^{\alpha}
+\frac{2\sqrt{2}}{3}\,\sum\limits_{\alpha\neq\beta}
\frac{\mathcal{D}_\alpha^\beta \chi^{iA}{}_\beta{}^\alpha}
{(\Delta^{+}_{\alpha\beta})^{1/2}\Delta^{-}_{\alpha\beta} }
\nonumber \\
&& +\frac{i2\sqrt{2}}{9}\,
\sum\limits_{\alpha,\beta,\gamma}
\frac{\chi^{A}_k{}_\alpha{}^\beta \Big( \chi^{(iB}{}_\beta{}^\gamma\chi{}^{k)}_B{}_\gamma{}^\alpha
+3\,\psi^{(i}{}_\beta{}^\gamma\bar\psi^{k)}{}_\gamma{}^\alpha
-3\,\bar\psi^{(i}{}_\beta{}^\gamma\psi^{k)}{}_\gamma{}^\alpha \Big)}
{(\Delta^{+}_{\alpha\beta})^{1/2} (\Delta{}_{\alpha\gamma}^{+})^{1/2}(\Delta^{+}_{\beta\gamma})^{1/2}}\,.
\label{Q-hidden-n}
\end{eqnarray}

\subsection{$\mathcal{N}{=}\,8$ superalgebras in the $n\,{=}\,1$ case}

\quad\, In the one-particle case the indices $\alpha,\beta$ take only one value and the harmonic variables are absent.
The Hamiltonian \p{Hamil} and the supercharges \p{Q-4-n}, \p{S-4-n}, \p{Q-hidden-n} in this case read
\be\label{H-1}
H = \frac{1}{2}\,p^2  + \frac{3 \,
\psi_{i}\psi^{i}\,\bar\psi_{k} \bar\psi^{k}
-12  \,\psi_{i}\bar\psi_{k}\chi^{iA}\chi^{k}_A
-\,\chi_{i}^{A}\chi^{}_{kA}\chi^{iB}\chi^{k}_B}{36\,x^{2}}\,,
\ee
\begin{eqnarray}\label{Q-1}
{Q}^i &=& p\, \psi^i- \frac{i\, \psi_{k}\,\big(2\,\psi^{(i}\bar\psi^{k)}
+3\,\chi^{iA} \chi^{k}_A \big)}{9\,x}\, ,\\
\label{Qb-1}
\bar{Q}_i &=& p\, \bar\psi_i+\frac{i\,\bar\psi^{k}\,\big(2\,\psi_{(i}\bar\psi_{k)}
+3\,\chi_i^{A} \chi_{kA} \big)}{9\,x}\,,
\end{eqnarray}
\be\label{S-1}
{S}^i = x\, \psi^i- t\,{Q}^i\, ,\qquad
\bar{S}_i = x\, \bar\psi_i - t\,\bar{Q}_i\,,
\ee
\be
\label{Qn-1}
\mathcal{Q}^{iA} = p\, \chi^{iA}+ \frac{i\, \chi^{A}_{k}\,\big(6\,\psi^{(i}\bar\psi^{k)}
+ \chi^{iB} \chi^{k}_B \big)}{9\,x}\,.
\ee

Using the $n\,{=}\,1$ case form of (\ref{CDB-2-nonh}),
\begin{equation}\label{CDB-2-nonh-1}
\{x, p\}^{**}= 1\,, \qquad
\{\psi{}^i, \bar\psi_k\}^{**}= -i\,\delta^i_k \,,\qquad
\{\chi^{iA}, \chi^{jB}\}^{**}= i\,\epsilon^{ij}\epsilon^{AB} \,,
\end{equation}
we arrive at the following Dirac brackets
for the fermionic generators \p{Q-1}, \p{Qb-1}, \p{S-1}
\begin{equation} \label{DB-QQ}
\begin{array}{c}
\{{Q}^{i}, \bar{Q}_{k}\}^{**}= -2i\delta^{i}_{k}H\,,\qquad  \{{Q}^{i}, {Q}^{k}\}^{**}=0
\,,\qquad  \{\bar{Q}_{i}, \bar{Q}_{k}\}^{**}=0\,, \\ [7pt]
\{{S}^{i}, \bar{S}_{k}\}^{**}= -2i\delta^{i}_{k}K\,,\qquad  \{{S}^{i}, {S}^{k}\}^{**}=0
\,,\qquad  \{\bar{S}_{i}, \bar{S}_{k}\}^{**}=0\,,
\\ [7pt]
\{{Q}^{i}, {S}^{k}\}^{**}= {\displaystyle \frac{4i}{3}}\,\epsilon^{ik}\,I^{(\psi)}\,,
\qquad
\{\bar{Q}_{i}, \bar{S}_{k}\}^{**}=-{\displaystyle \frac{4i}{3}}\,\epsilon_{ik}\,\bar I^{(\psi)}\,,
\\ [7pt]
\{{Q}^{i}, \bar{S}_{k}\}^{**}= 2i\delta^{i}_{k}\,D-{\displaystyle \frac{2i}{3}}\, J^i{}_k-{\displaystyle \frac{4i}{3}}\,\delta^{i}_{k}\,I^{(\psi)}_3\,,
\\ [7pt]
\{\bar{Q}_{i}, {S}^{k}\}^{**}=2i\delta_{i}^{k}\,D+{\displaystyle \frac{2i}{3}}\, J_i{}^k+{\displaystyle \frac{4i}{3}}\,\delta_{i}^{k}\,I^{(\psi)}_3\,.
\end{array}
\end{equation}
Here, the bosonic generators $H$ (defined in \p{Hamil}) and the generators
\begin{equation}\label{KD-cl}
K ={\textstyle\frac{1}{2}}\,x^2  -t\,x p +    t^2\, H\,,\qquad
D =-{\textstyle\frac{1}{2}}\,x p + t\, H\,,
\end{equation}
\begin{equation} \label{I-cl}
I^{(\psi)} = {\textstyle\frac{i}{2}}\, \psi_k\psi^k\,,\qquad
\bar I^{(\psi)} = {\textstyle\frac{i}{2}}\, \bar\psi_k\bar\psi^k\,,\qquad
I^{(\psi)}_3 ={\textstyle\frac{i}{2}}\,  \psi_k\bar\psi^k
\end{equation}
\begin{equation} \label{T-cl}
J_{ik} = -i\left[ \psi_{(i}\bar\psi{}_{k)}-{\textstyle\frac12}\,\chi_i^{A}\chi{}_{kA}\right],
\end{equation}
form the following algebra
\begin{equation}\label{DB-T-1}
\{H, K\}^{**}= 2D\,,\qquad
\{H, D\}^{**}= H\,,\qquad
\{K, D\}^{**}= -K\,,
\end{equation}
\begin{equation}\label{DB-I}
\{I^{(\psi)}, \bar I^{(\psi)}\}^{**}= 2I^{(\psi)}_3\,,\qquad
\{I^{(\psi)}, I^{(\psi)}_3\}^{**}= I^{(\psi)} \,,\qquad
\{\bar I^{(\psi)}, I^{(\psi)}_3\}^{**}= -\bar I^{(\psi)}\,,\qquad
\end{equation}
\begin{equation}\label{DB-J}
\{J_{ij}, J_{kl}\}^{**}=\epsilon_{ik}J_{jl} +\epsilon_{jl}J_{ik}\,.
\end{equation}

Finally, defining the quantities ${Q}_{\mu i i^\prime}$, $T_{\mu\nu}$, $I^{(\psi)}_{i^\prime k^\prime}$ as
\begin{equation}\label{not-Q}
{Q}_{1 i1^\prime}={Q}_{i}\,,\qquad {Q}_{1 i2^\prime}=\bar{Q}_{i}\,,
\qquad\qquad
{Q}_{2 i1^\prime}=-{S}_{i}\,,\qquad {Q}_{2 i2^\prime}=-\bar{S}_{i}\,,
\end{equation}
\begin{equation}\label{not-T}
T_{11}=H\,,\qquad T_{22}=K\,,\qquad T_{12}=D\,,
\end{equation}
\begin{equation} \label{I-cl1}
I^{(\psi)}_{1^\prime 1^\prime} = I^{(\psi)}\,,\qquad
I^{(\psi)}_{2^\prime 2^\prime} = \bar I^{(\psi)}\,,\qquad
I^{(\psi)}_{1^\prime 2^\prime} =I^{(\psi)}_3 \,,
\end{equation}
we find the closed superalgebra of the full set of generators:
\begin{equation} \label{DB-Q-g}
\{{Q}_{\mu i i^\prime}, {Q}_{\nu k k^\prime}\}^{**}= -2i\Big(\epsilon_{ik}\epsilon_{i^\prime
k^\prime} T_{\mu\nu}-\frac13\,\epsilon_{\mu\nu}\epsilon_{i^\prime k^\prime} J_{ik}-\frac23\,
\epsilon_{\mu\nu}\epsilon_{ik} I^{(\psi)}_{i^\prime k^\prime}\Big)\,,
\end{equation}
\begin{equation}\label{DB-T}
\{T_{\mu\nu}, T_{\lambda\rho}\}^{**}= \epsilon_{\mu\lambda}T_{\nu\rho} +\epsilon_{\nu\rho}T_{\mu\lambda}\,,
\end{equation}
\begin{equation}\label{DB-I-c}
\{I^{(\psi)}_{i^\prime j^\prime}, I^{(\psi)}_{k^\prime l^\prime}\}^{**}= \epsilon_{i^\prime k^\prime}I^{(\psi)}_{j^\prime l^\prime}
+\epsilon_{j^\prime l^\prime}I^{(\psi)}_{i^\prime k^\prime}\,,
\end{equation}
\begin{equation}\label{DB-JQ}
\begin{array}{rcl}
\{T_{\mu\nu}, {Q}_{\lambda i i^\prime}\}^{**}&=&-\epsilon_{\lambda(\mu}{Q}_{\nu) i i^\prime} \,, \\ [7pt]
\{J_{ij},{Q}_{\mu k i^\prime}\}^{**}&=&-\epsilon_{k(i}{Q}_{\mu i^\prime j)}\,, \\ [7pt]
\{I^{(\psi)}_{i^\prime j^\prime},{Q}_{\mu i k^\prime}\}^{**}&=&-\epsilon_{k^\prime (i^\prime}{Q}_{\mu j^\prime ) i}\,.
\end{array}
\end{equation}
This is none other than the standard form of the superalgebra $D(2,1;\alpha{=}{-}1/3)$.

In terms of the quantities
\begin{equation} \label{T-cl-2}
J^{(\psi)}_{ik} = -i \psi_{(i}\bar\psi{}_{k)}\,,\qquad J^{(\chi)}_{ik}=\frac{i}{2} \,\chi_i^{A}\chi_{kA} \,,
\end{equation}
the ``currents'' \p{T-cl} are written as
\begin{equation} \label{T-cl-1}
J^{ik} = J_{(\psi)}^{ik}+J_{(\chi)}^{ik}\,.
\end{equation}
Using the definition \p{T-cl-2}, we can represent the Hamiltonian \p{H-1} and the supercharges \p{Q-1}, \p{Qb-1} in the form
\begin{eqnarray}\label{H-1a}
H &=& \frac{1}{2}\,p^2  + \frac{J^{(\psi)}{}^{ik}J^{(\psi)}_{ik}+J^{(\chi)}{}^{ik}J^{(\chi)}_{ik}
-6J^{(\psi)}{}^{ik}J^{(\chi)}{}_{ik}}{9\,x^{2}}\,, \\
\label{Q-1a}
{Q}^i &=& p\, \psi^i+\frac{2}{9x}\, \psi_{k}\,\big( J^{(\psi)}{}^{ik}-3J^{(\chi)}{}^{ik}\big) \, ,\\
\label{Qb-1a}
\bar{Q}_i &=& p\, \bar\psi_i-\frac{2}{9x}\, \bar\psi^{k}\,\big( J^{(\psi)}_{ik}-3J^{(\chi)}_{ik}\big)\,.
\end{eqnarray}
After passing to the notation
\begin{equation}\label{spinors-chi}
\chi_{k}=i\chi_{k,A=1}\,,\qquad \bar\chi_{i}=i\chi_{k,A=2}\,,\qquad  (\chi^{i})^*=\bar\chi_{i} \,,
\end{equation}
the ``current'' $J^{(\chi)}_{ik}$ can be cast in the form similar to $J^{(\psi)}_{ik}$,
\begin{equation} \label{T-cl-2-chi}
J^{(\chi)}_{ik} = -i \chi_{(i}\bar\chi{}_{k)}\,,
\end{equation}
and the supercharges \p{Qn-1} become
\begin{eqnarray}
\label{Q-n-1a}
\mathcal{Q}^i &=& p\, \chi^i+\frac{2}{9x}\, \chi_{k}\,\big( J_{(\chi)}^{ik}-3J_{(\psi)}^{ik}\big) \, ,\\
\label{Qb-n-1a}
\bar{\mathcal{Q}}_i &=& p\, \bar\chi_i-\frac{2}{9x}\, \bar\chi^{k}\,\big( J_{(\chi)}{}_{ik}-3J_{(\psi)}{}_{ik}\big)\,,
\end{eqnarray}
where $\mathcal{Q}_k=i\mathcal{Q}_{k,A=1}$, $\bar{\mathcal{Q}}_k=i\mathcal{Q}_{k,A=2}$.

The supercharges \p{Q-n-1a}, \p{Qb-n-1a} are obtained from
the supercharges \p{Q-1a}, \p{Qb-1a} via the substitutions $\chi^i \leftrightarrow\psi^i$, $\bar\chi_i \leftrightarrow\bar\psi_i$.
Dirac brackets of the supercharges \p{Q-n-1a}, \p{Qb-n-1a} with $T_{\mu\nu}$ yield the additional fermionic charges
\begin{equation}\label{S-d}
\mathcal{S}^{i} = x\, \chi^{i}- t\,\mathcal{Q}^{i}\,,\qquad  \bar{\mathcal{S}}_{i} = x\, \bar\chi_{i}- t\,\bar{\mathcal{Q}}_{i}\,.
\end{equation}
Defining, similarly to \p{not-Q}, the quantities ${Q}_{\mu i A}$ by
\begin{equation}\label{not-Q1}
{\mathcal{Q}}_{1 i,A=1}={\mathcal{Q}}_{i}\,,\qquad {\mathcal{Q}}_{1 i,A=2}=\bar{\mathcal{Q}}_{i}\,,
\qquad
{\mathcal{Q}}_{2 i,A=1}=-{\mathcal{S}}_{i}\,,\qquad {\mathcal{Q}}_{2 i,A=2}=-\bar{\mathcal{S}}_{i}\,,
\end{equation}
we obtain Dirac brackets which are analogous to \p{DB-Q-g}
\begin{equation} \label{DB-Q-g1}
\{{\mathcal{Q}}_{\mu i A}, {\mathcal{Q}}_{\nu k B}\}^{**}= -2i\Big(\epsilon_{ik}\epsilon_{A
B} T_{\mu\nu}-\frac13\,\epsilon_{\mu\nu}\epsilon_{AB} J_{ik}-\frac23\,
\epsilon_{\mu\nu}\epsilon_{ik} I^{(\chi)}_{AB}\Big)\,.
\end{equation}
Therefore, the supercharges \p{Q-n-1a}, \p{Qb-n-1a} and \p{S-d} also constitute the superalgebra \, $D(2,1;\alpha{=}{-}1/3)$,
with the bosonic generators $T^{\mu\nu}$, $J_{ij}$ defined by the same expressions \p{Q-1a}, \p{Qb-1a} as in
the previously discussed $D(2,1;\alpha{=}{-}1/3)$ superalgebra,
and with the generators
\begin{equation} \label{I-cl-chi}
I^{(\chi)} = {\textstyle\frac{i}{2}}\, \chi_k\chi^k\,,\qquad
\bar I^{(\chi)} = {\textstyle\frac{i}{2}}\, \bar\chi_k\bar\chi^k\,,\qquad
I^{(\chi)}_3 ={\textstyle\frac{i}{2}}\,  \chi_k\bar\chi^k
\end{equation}
instead of  $I^{(\psi)}$, $\bar I^{(\psi)}$, $I^{(\psi)}_3$.
The quantities defined similarly to \p{I-cl1},
\begin{equation} \label{I-cl2}
I^{(\chi)}_{11} = I^{(\chi)}\,,\qquad
I^{(\chi)}_{22} = \bar I^{(\chi)}\,,\qquad
I^{(\chi)}_{12} =I^{(\chi)}_3,
\end{equation}
are combined into the generators
\begin{equation} \label{T-cl-3}
I^{(\chi)}_{AB}=\frac{i}{2} \,\chi_A^{k}\chi_{kB},
\end{equation}
which satisfy the relations
\begin{equation}\label{DB-I-c2}
\{I^{(\chi)}_{AB}, I^{(\chi)}_{CD}\}^{**}= \epsilon_{AC}I^{(\chi)}_{BD}
+\epsilon_{BD}I^{(\chi)}_{AC}\,.
\end{equation}

The crossing Dirac brackets among the supercharges of the two $\mathcal{N}{=}\,4$ supersymmetries are vanishing:
\begin{equation}\label{DB-Q1-Q2}
\{{Q}_{i}, \mathcal{Q}_{j}\}^{**}=  \{{Q}_{i}, \bar{\mathcal{Q}}_{j}\}^{**}=
\{\bar{Q}_{i}, \mathcal{Q}_{j}\}^{**}=  \{\bar{Q}_{i}, \bar{\mathcal{Q}}_{j}\}^{**}=0\,.
\end{equation}
Also,
\begin{equation}\label{DB-S1-S2}
\{{S}_{i}, \mathcal{S}_{j}\}^{**}=  \{{S}_{i}, \bar{\mathcal{S}}_{j}\}^{**}=
\{\bar{S}_{i}, \mathcal{S}_{j}\}^{**}=  \{\bar{S}_{i}, \bar{\mathcal{S}}_{j}\}^{**}=0\,.
\end{equation}
The only non-vanishing Dirac brackets are
\begin{eqnarray}\label{DB-Q1-S2}
&&\{{Q}_{i}, \mathcal{S}_{j}\}^{**}=  -\{\mathcal{Q}_{i}, {S}_{j}\}^{**}=-\frac43\,N_{ij}\,,\quad
\{\bar{Q}_{i}, \bar{\mathcal{S}}_{j}\}^{**}=  -\{\bar{\mathcal{Q}}_{i}, \bar{S}_{j}\}^{**}=-\frac43\,\bar N_{ij}\,,\\
&&\{{Q}_{i}, \bar{\mathcal{S}}_{j}\}^{**}=  -\{\bar{\mathcal{Q}}_{i}, {S}_{j}\}^{**}=-\frac43\,K_{ij}\,,\quad
\{\bar{Q}_{i}, {\mathcal{S}}_{j}\}^{**}=  -\{{\mathcal{Q}}_{i}, \bar{S}_{j}\}^{**}=-\frac43\,\bar K_{ij}\,,
\label{DB-S1-Q2}
\end{eqnarray}
where
\begin{equation}\label{N-def}
N_{ij}:=\psi_{(i}\chi_{j)}\,,\quad \bar N_{ij}:=\bar\psi_{(i}\bar\chi_{j)}\,,\qquad
K_{ij}:=\psi_{(i}\bar\chi_{j)}\,,\quad \bar K_{ij}:=\bar\psi_{(i} \chi_{j)} \,.
\end{equation}
Upon using the three-spinor supercharges \p{not-Q}, \p{not-Q1}, the anticommutators \p{DB-Q1-Q2}, \p{DB-S1-S2}, \p{DB-Q1-S2}, \p{DB-S1-Q2}
can be succinctly written as
\begin{equation} \label{DB-QS-g1}
\{Q_{\mu i i^\prime}, {\mathcal{Q}}_{\nu k B}\}^{**}= \frac43\, N_{(ik)i^\prime A}\,,
\end{equation}
where
\begin{equation}\label{N-3}
N_{(ik)1^\prime 1}=N_{ik}\,, \qquad
N_{(ik)2^\prime 2}=\bar N_{ik}\,,\qquad
N_{(ik)1^\prime 2}=K_{ik}\,, \qquad
N_{(ik)2^\prime 1}=\bar K_{ik}\,.
\end{equation}
Using \p{spinors-chi} and introducing the second two-rank spinor $\psi_{kk^\prime}$ as
\begin{equation}\label{spinors-psi}
\psi_{k}=i\psi_{k,k^\prime=1^\prime}\,,\qquad \bar\psi_{i}=i\psi_{k,k^\prime=2^\prime}\,,\qquad  (\psi^{ii^\prime})^*=-\psi_{ii^\prime} \,,
\end{equation}
we rewrite the generators \p{N-def}, \p{N-3} in the form
\begin{equation}\label{N-3-g}
N_{(ij)k^\prime A}=-\psi_{(i}{}_{k^\prime}\chi_{j)A}\,.
\end{equation}
Also,  in terms of the quantities \p{spinors-psi} the $su(2)$ generators \p{I-cl}, \p{I-cl1} become
\begin{equation} \label{T-cl-4}
I^{(\psi)}_{i^\prime j^\prime}=\frac{i}{2} \,\psi_{i^\prime}^{k}\psi_{k j^\prime}\,.
\end{equation}

The bosonic generators $T_{\mu\nu}$ form $sl(2,\mathbb{R})$ algebra.
The generators $N_{(ij)k^\prime A}$ refer to the
${\displaystyle{\frac{\mathrm{SO}(7)}{\mathrm{SU}(2)\otimes \mathrm{SU}(2)\otimes \mathrm{SU}(2)}}}$ coset and, together with
the $su(2)\oplus su(2)\oplus su(2)$ generators $J_{ij}$, $I^{(\psi)}_{i^\prime j^\prime}$, $I^{(\chi)}_{AB}$, form just $so(7)$ R-symmetry algebra
\footnote{Such a decomposition of $so(7)$ algebra was used in \cite{BMW-1982,JW-1984}.}
\begin{equation}\label{s7-str}
so(7):\qquad\quad J_{ij}\qquad I^{(\psi)}_{i^\prime j^\prime}\qquad I^{(\chi)}_{AB}\qquad N_{(ij)\,k^\prime A} \,.
\end{equation}
These bosonic generators and $T_{\mu\nu}$, together with the odd generators $Q_{\mu i i^\prime}$, $\mathcal{Q}_{\mu i A}$, constitute $F(4)$ superalgebra.
We observe that the $F(4)$ superalgebra obtained in this way has the following notable structure
\begin{equation}\label{F4-str}
F(4):\qquad\quad \lefteqn{\overbrace
{\phantom{\,\, I^{(\psi)}_{i^\prime j^\prime}\quad Q_{\mu i i^\prime}\quad T_{\mu\nu}\quad J_{ij} \,\,}}^{D(2,1;\alpha{=}{-}1/3)}}
\,\, I^{(\psi)}_{i^\prime j^\prime}\quad Q_{\mu i i^\prime}\,\,\,
\underbrace{\,\,\,T_{\mu\nu}\quad J_{ij}\quad \mathcal{Q}_{\mu i A}\quad I^{(\chi)}_{AB}\,\,}_{D(2,1;\alpha{=}{-}1/3)}
\,\,\,N_{(ij)\,k^\prime A} \,.
\end{equation}
It includes two $D(2,1;\alpha\,{=}{-}1/3)$ superalgebras with the common $sl(2,\mathbb{R})$ and $su(2)$ generators
$T_{\mu\nu}$ and $J_{ij}$ and so can be treated as a closure of these two superalgebras.

\subsection{$\mathcal{N}{=}\,8$ superalgebras for $n\,{>}\,1$}

\quad\, Let us introduce, similarly to \p{spinors-chi}, the quantities
\begin{equation}\label{spinors-chi-n}
\chi_{k}{}_\alpha{}^\beta=i\chi_{k,A=1,}{}_\alpha{}^\beta\,,\qquad
\bar\chi_{k}{}_\alpha{}^\beta=i\chi_{k,A=2,}{}_\alpha{}^\beta\,,
\end{equation}
which satisfy the relations
\begin{equation}\label{CDB-2-chi-1}
(\chi^{i}{}_\alpha{}^\beta)^*=\bar\chi_{i}{}_\beta{}^\alpha \,,\qquad
\{\chi{}^i{}_\alpha{}^\beta, \bar\chi_k{}_\gamma{}^\delta\}^{**}= -i\,\delta^i_k \delta^\delta_\alpha \delta^\beta_\gamma\,.
\end{equation}
In terms of the newly defined objects  the supercharges \p{Q-4-n} take the form
\begin{eqnarray}\label{Q-4-n1}
\!\!\!\!\!\!\!\!\!\!   Q^i \!\! &=&\!\!  \sum\limits_{\alpha} p_\alpha \psi^i{}_{\alpha}{}^{\alpha}
+\frac{2\sqrt{2}}{3}\,\sum\limits_{\alpha\neq\beta}
\frac{\mathcal{D}_\alpha^\beta \psi{}^i{}_\beta{}^\alpha}
{(\Delta^{+}_{\alpha\beta})^{1/2}\Delta^{-}_{\alpha\beta} }
\\
&&\!\! -\frac{i2\sqrt{2}}{9}\,
\sum\limits_{\alpha,\beta,\gamma}
\frac{\psi_k{}_\alpha{}^\beta \Big(\psi^{(i}{}_\beta{}^\gamma\bar\psi^{k)}{}_\gamma{}^\alpha
-\bar\psi^{(i}{}_\beta{}^\gamma\psi^{k)}{}_\gamma{}^\alpha-3
\chi^{(i}{}_\beta{}^\gamma\bar\chi^{k)}{}_\gamma{}^\alpha
+3\bar\chi^{(i}{}_\beta{}^\gamma\chi^{k)}{}_\gamma{}^\alpha \Big)}
{(\Delta^{+}_{\alpha\beta})^{1/2} (\Delta{}_{\alpha\gamma}^{+})^{1/2}(\Delta^{+}_{\beta\gamma})^{1/2}}\,,
\nonumber \\
\!\!\!\!\!\!\!\!\!\!    \bar Q_i &=& \sum\limits_{\alpha} p_\alpha \bar\psi_i{}_{\alpha}{}^{\alpha}
+\frac{2\sqrt{2}}{3}\,\sum\limits_{\alpha\neq\beta}
\frac{\mathcal{D}_\alpha^\beta \bar\psi{}_i{}_\beta{}^\alpha}
{(\Delta^{+}_{\alpha\beta})^{1/2}\Delta^{-}_{\alpha\beta} }
\label{Q-4-n12} \\
&&\!\! +\frac{i2\sqrt{2}}{9}\,
\sum\limits_{\alpha,\beta,\gamma}
\frac{\bar\psi^k{}_\alpha{}^\beta \Big(\psi_{(i}{}_\beta{}^\gamma\bar\psi_{k)}{}_\gamma{}^\alpha
-\bar\psi_{(i}{}_\beta{}^\gamma\psi_{k)}{}_\gamma{}^\alpha
-3\chi_{(i}{}_\beta{}^\gamma\bar\chi_{k)}{}_\gamma{}^\alpha
+3\bar\chi_{(i}{}_\beta{}^\gamma\chi_{k)}{}_\gamma{}^\alpha \Big)}
{(\Delta^{+}_{\alpha\beta})^{1/2} (\Delta{}_{\alpha\gamma}^{+})^{1/2}(\Delta^{+}_{\beta\gamma})^{1/2}}\,.
\nonumber
\end{eqnarray}
The supercharges \p{Q-hidden-n} are given by the analogous  expressions
\begin{eqnarray}\label{Q-hidden-n1}
\!\!\!\!\!\!\!\!\!\!   \mathcal{Q}^i \!\! &=&\!\!  \sum\limits_{\alpha} p_\alpha \chi^i{}_{\alpha}{}^{\alpha}
+\frac{2\sqrt{2}}{3}\,\sum\limits_{\alpha\neq\beta}
\frac{\mathcal{D}_\alpha^\beta \chi{}^i{}_\beta{}^\alpha}
{(\Delta^{+}_{\alpha\beta})^{1/2}\Delta^{-}_{\alpha\beta} }
\\
&&\!\! -\frac{i2\sqrt{2}}{9}\,
\sum\limits_{\alpha,\beta,\gamma}
\frac{\chi_k{}_\alpha{}^\beta \Big(\chi^{(i}{}_\beta{}^\gamma\bar\chi^{k)}{}_\gamma{}^\alpha
-\bar\chi^{(i}{}_\beta{}^\gamma\chi^{k)}{}_\gamma{}^\alpha-3
\psi^{(i}{}_\beta{}^\gamma\bar\psi^{k)}{}_\gamma{}^\alpha
+3\bar\psi^{(i}{}_\beta{}^\gamma\psi^{k)}{}_\gamma{}^\alpha \Big)}
{(\Delta^{+}_{\alpha\beta})^{1/2} (\Delta{}_{\alpha\gamma}^{+})^{1/2}(\Delta^{+}_{\beta\gamma})^{1/2}}\,,
\nonumber \\
\!\!\!\!\!\!\!\!\!\!    \bar {\mathcal{Q}}_i &=& \sum\limits_{\alpha} p_\alpha \bar\chi_i{}_{\alpha}{}^{\alpha}
+\frac{2\sqrt{2}}{3}\,\sum\limits_{\alpha\neq\beta}
\frac{\mathcal{D}_\alpha^\beta \bar\chi{}_i{}_\beta{}^\alpha}
{(\Delta^{+}_{\alpha\beta})^{1/2}\Delta^{-}_{\alpha\beta} }
\label{Q-hidden-n2} \\
&&\!\! +\frac{i2\sqrt{2}}{9}\,
\sum\limits_{\alpha,\beta,\gamma}
\frac{\bar\chi^k{}_\alpha{}^\beta \Big(\chi_{(i}{}_\beta{}^\gamma\bar\chi_{k)}{}_\gamma{}^\alpha
-\bar\chi_{(i}{}_\beta{}^\gamma\chi_{k)}{}_\gamma{}^\alpha
-3\psi_{(i}{}_\beta{}^\gamma\bar\psi_{k)}{}_\gamma{}^\alpha
+3\bar\psi_{(i}{}_\beta{}^\gamma\psi_{k)}{}_\gamma{}^\alpha \Big)}
{(\Delta^{+}_{\alpha\beta})^{1/2} (\Delta{}_{\alpha\gamma}^{+})^{1/2}(\Delta^{+}_{\beta\gamma})^{1/2}}\,.
\nonumber
\end{eqnarray}
The supercharges \p{Q-4-n1}, \p{Q-4-n12} and \p{Q-hidden-n1}, \p{Q-hidden-n2} go into each other through the replacements
\begin{equation}\label{spinors-repl}
(\psi_{k}{}_\alpha{}^\beta, \bar\psi_{k}{}_\alpha{}^\beta)\quad \leftrightarrow\quad
(\chi_{k}{}_\alpha{}^\beta, \bar\chi_{k}{}_\alpha{}^\beta)\,.
\end{equation}

The supercharges \p{Q-4-n1}, \p{Q-4-n12} and \p{Q-hidden-n1}, \p{Q-hidden-n2}, together with the generators \p{S-4-n}, produce
the conformal $F(4)$ superalgebra considered in the previous subsection.
The Hamiltonian in the multiparticle case is given in \p{Hamil}.
The remaining even generators have the same form as in  \p{I-cl1}, \p{DB-I-c2}, \p{T-cl-1}, \p{T-cl-2}, \p{N-def},
with the matrix variables and the traces of their products.
For example, one of the involved quantities is
\begin{equation} \label{T-cl-2-matr}
J_{ik} = -i \mathrm{Tr} \left( \psi_{(i}\bar\psi{}_{k)}+\chi_{(i}\bar\chi{}_{k)} \right) .
\end{equation}
The basic difference of the multiparticle ($n\,{>}\,1$) case  from its one-particle ($n\,{=}\,1$) prototype
lies in the property that the corresponding fermionic charges form a closed algebra in a weak sense, that is,
on the shell of the first class constraints \p{constr-1}; for example,
\begin{equation}\label{Q-Q-5}
\{Q^i,Q^j\}=-\frac89\sum\limits_{\alpha\neq\beta}
\frac{\psi{}^i{}_\alpha{}^\beta\psi{}^j{}_\beta{}^\alpha}
{\Delta^{+}_{\alpha\beta}(\Delta^{-}_{\alpha\beta})^{2} }\left(\mathcal{D}_\alpha-\mathcal{D}_\beta \right)\,.
\end{equation}
Such terms arise from the Dirac brackets $\{\mathcal{D}_\alpha^\beta,\mathcal{D}_\gamma^\delta\}$
of the quantities $\mathcal{D}_\alpha^\beta$ which are present in the supercharges  \p{Q-4-n1}, \p{Q-4-n12}, \p{Q-hidden-n1}, \p{Q-hidden-n2}.
A similar situation occurred in the matrix model of refs.\,\cite{FI,FILS}.

It is worth noting that the second terms in the ${\cal N}{=}\,4$ supercharges \p{Q-4-n1}, \p{Q-4-n12}
involve off-diagonal fermionic fields $\psi{}^i{}_\beta{}^{\alpha}$, $\bar\psi{}_i{}_\beta{}^{\alpha}$, $\alpha\,{\neq}\, \beta$.
Although the relevant sums also contain the diagonal fermionic fields $\psi{}^i{}_\alpha{}^{\alpha}$, $\bar\psi{}_i{}_\alpha{}^{\alpha}$,
in these sums there are some terms  which do not contain diagonal fermionic fields at all.
The matrix model of ref.\,\cite{FILS} (see also \cite{KLS-18}) reveals a similar structure of the supercharges.
This is an essential difference from the non-matrix models (see, e.g., \cite{Gal-2017,GLP-2007}),
in which {\it all} terms in the supercharges contain diagonal-like fermionic fields
(same as in the terms with the canonical momenta).

\section{Concluding remarks}

\quad\, In this paper we have presented multiparticle ${\cal N}{=}\,8$ superconformal mechanics
with the underlying  $F(4)$ supersymmetry.
The initial system in our construction is the ${\cal N}{=}\,4$ superfield system with
the matrix bosonic $({\bf 1, 4, 3})$ and the matrix fermionic $({\bf 0, 4, 4})$ multiplets.
This system exhibits an implicit ${\cal N}{=}\,4$ supersymmetry that extends  ${\cal N}{=}\,4$
superconformal symmetry $D(2,1;\alpha\,{=}{-}1/3)$ to ${\cal N}{=}\,8$ superconformal symmetry  $F(4)$
with respect to which the ${\cal N}=4$ multiplets are combined into irreducible $({\bf 1, 8, 7})$ supermultiplets.

The off-diagonal bosonic components of matrix $({\bf 1, 4, 3})$ multiplet take values in the flag coset manifold $\mathrm{U}(n)/[\mathrm{U}(1)]^n$
and can be identified with a kind of the target space harmonics.
We presented the full set of the first class constraints generating gauge transformations of these variables.

In our model the harmonic variables are dynamical -- the Lagrangian is of the second order in the harmonic velocities.
Moreover, the covariant harmonic momenta are present in the Calogero-like terms in the supercharges
(the second terms in  \p{Q-4-n1}, \p{Q-hidden-n1}) and in the Hamiltonian
(the second term in  \p{Hamil}).
It would be  interesting to inquire whether some ${\cal N}{=}\,8$ supersymmetric reduction in the harmonic sector is possible,
such that the  numerator(s) in the second terms of the Hamiltonian  \p{Hamil} become constants and the latter acquires  a
more Calogero-like form.

One more interesting problem for the future study is to find a quantum realization of the ${\cal N}{=}\,8$ superalgebras we have discussed.

Let us emphasize once more that in this article we dealt with the case when all eigenvalues of the matrix $M$  \p{M-UY} are unequal,
$y_\alpha\,{\neq}\, y_\beta$ for any $\alpha\,{\neq}\, \beta$. This case is most general, in the sense that all options
with coincident eigenvalues can be recovered as some special limits of it.
On the other hand, the cases with equal eigenvalues could be equally considered on their own right along the same lines as here.
The basic new point will be the appearance of the equalities $\Delta^{-}_{\alpha\beta}\,{=}\, 0$ when $y_\alpha\,{=}\, y_\beta$.
This will amount to additional first-class constraints apart from the $\mathrm{U}(1)$-constraints \p{constr-1}.
If $k_1$ eigenvalues coincide with each other, there  will arise the constraints that generate local symmetry $\mathrm{U}(k_1)$.
In general,  in the system with $y_{\alpha_{1}}\,{=}\,y_{\alpha_{2}}\,{=}\,\ldots \,{=}\,y_{\alpha_{k_1}}$,
$y_{\beta_{1}}\,{=}\,y_{\beta_{2}}\,{=}\,\ldots \,{=}\,y_{\beta_{k_2}}$, {\it etc.},
the relevant harmonics parametrize the coset\footnote{Such a coset is in fact a most general flag manifold of SU(n), while the coset
which we dealt with in this paper is the maximal-dimension flag manifold containing all others as submanifolds.}
$\,{\displaystyle \frac{\mathrm{U}(n)}{\mathrm{U}(k_1){\otimes} \mathrm{U}(k_2){\otimes}\ldots{\otimes} \mathrm{U}(k_m)}}$,
$k_1\,{+}\,k_2\,{+}\,\ldots \,{+}\,k_m=n$.
We plan to investigate multiparticle models with such harmonic sets elsewhere.

\bigskip
\section*{\quad \ \ Acknowledgements}
This work was supported by the Russian Science Foundation, grant No.\,16-12-10306. E.I. thanks Laboratoire de Physique, ENS-Lyon, for the kind hospitality
at the last stage of this study.

\bigskip
\section*{\quad \ \ Appendix: Calogero-like form of the supercharges \\ ${}$ \hspace{4.7cm} in non-flat target space}
\def\theequation{A.\arabic{equation}}
\setcounter{equation}0

\quad\, In terms of the odd variables $\underset{{}^{{}^\sim}}{\varphi}{}^i{}_\alpha{}^\beta$,
$\underset{{}^{{}^\sim}}{\bar\varphi}{}^i{}_\alpha{}^\beta$,
$\underset{{}^{{}^\sim}}{\phi}{}^i{}_\alpha{}^\beta$, $\underset{{}^{{}^\sim}}{\bar\phi}{}^i{}_\alpha{}^\beta$
defined in \p{new-ferni-matr}
[similarly to \p{spinors-chi-n}, $\underset{{}^{{}^\sim}}{\phi}{}^i{}_\alpha{}^\beta$,
$\underset{{}^{{}^\sim}}{\bar\phi}{}^i{}_\alpha{}^\beta$
are the components of $\underset{{}^{{}^\sim}}{\phi}{}^{iA}_\alpha{}^\beta$],
the supercharges \p{Q-4-n1} and \p{Q-hidden-n1} take the form
\begin{eqnarray}\label{Q-4-n1-ap}
\!\!\!\!\!\!\!\!\!\!   Q^i \!\! &=&\!\!  \sum\limits_{\alpha} p^{(y)}_\alpha \underset{{}^{{}^\sim}}{\varphi}{}^i{}_{\alpha}{}^{\alpha}
+ \sum\limits_{\alpha\neq\beta} (y_{\alpha} - y_{\beta})^{-1}
\mathcal{D}_\alpha^\beta \underset{{}^{{}^\sim}}{\varphi}{}^i{}_\beta{}^\alpha
-\frac{i}{3} \sum\limits_{\alpha,\beta} \underset{{}^{{}^\sim}}{\varphi}{}_k{}_\alpha{}^\beta \Big(
[\underset{{}^{{}^\sim}}{\varphi}{}^{(i}, \underset{{}^{{}^\sim}}{\bar\varphi}{}^{k)}]
-3[\underset{{}^{{}^\sim}}{\phi}{}^{(i}, \underset{{}^{{}^\sim}}{\phi}{}^{k)}]
\Big){}_\beta{}^\alpha
,
\\
\!\!\!\!\!\!\!\!\!\!   \mathcal{Q}^i \!\! &=&\!\!  \sum\limits_{\alpha} p^{(y)}_\alpha \underset{{}^{{}^\sim}}{\phi}{}^i{}_{\alpha}{}^{\alpha}
+ \sum\limits_{\alpha\neq\beta} (y_{\alpha} - y_{\beta})^{-1}
\mathcal{D}_\alpha^\beta \underset{{}^{{}^\sim}}{\phi}{}^i{}_\beta{}^\alpha
-\frac{i}{3} \sum\limits_{\alpha,\beta} \underset{{}^{{}^\sim}}{\phi}{}_k{}_\alpha{}^\beta \Big(
[\underset{{}^{{}^\sim}}{\phi}{}^{(i}, \underset{{}^{{}^\sim}}{\bar\phi}{}^{k)}]
-3[\underset{{}^{{}^\sim}}{\varphi}{}^{(i}, \underset{{}^{{}^\sim}}{\varphi}{}^{k)}]
\Big){}_\beta{}^\alpha \label{Q-4-bn1-ap}
\end{eqnarray}
(and c.c.). The supercharges \p{Q-4-n1-ap} \p{Q-4-bn1-ap} act in the phase space
parametrized by the bosonic variables $y_\alpha$ defined in  \p{Y-matrix}, their momenta $p^{(y)}_\alpha$ and harmonic variables.

The supercharges \p{Q-4-n1-ap}, \p{Q-4-bn1-ap} have the Calogero-like form, as is seen from their second
terms involving the typical denominators $({\displaystyle y_{\alpha} - y_{\beta}})\,$.
But, as opposed to the ``flat'' Dirac brackets \p{CDB-2-nonh}, Dirac brackets for the ``undertilded'' variables used here
have extra non-vanishing terms: the full set of the non-vanishing Dirac brackets is encompassed by the relations
\begin{equation}\label{CDB-2-1ap}
\{y_\alpha, p^{(y)}_\beta\}^{**}= \delta_{\alpha\beta}\,,
\end{equation}
\begin{equation}\label{CDB-2-2ap}
\{\underset{{}^{{}^\sim}}{\varphi}{}^i{}_\alpha{}^\beta, p^{(y)}_\gamma\}^{**}=
-\frac{\delta_{\alpha\gamma} + \delta_{\beta\gamma}}{2(y_\alpha+y_\beta)}\,\underset{{}^{{}^\sim}}{\varphi}{}^i{}_\alpha{}^\beta\,,
\qquad
\{\underset{{}^{{}^\sim}}{\bar\varphi}{}_i{}_\alpha{}^\beta, p^{(y)}_\gamma\}^{**}=
-\frac{\delta_{\alpha\gamma} + \delta_{\beta\gamma}}{2(y_\alpha+y_\beta)}\,\underset{{}^{{}^\sim}}{\bar\varphi}{}_i{}_\alpha{}^\beta\,,
\end{equation}
\begin{equation}\label{CDB-2-3ap}
\{\underset{{}^{{}^\sim}}{\phi}{}^i{}_\alpha{}^\beta, p^{(y)}_\gamma\}^{**}=
-\frac{\delta_{\alpha\gamma} + \delta_{\beta\gamma}}{2(y_\alpha+y_\beta)}\,\underset{{}^{{}^\sim}}{\phi}{}^i{}_\alpha{}^\beta\,,
\qquad
\{\underset{{}^{{}^\sim}}{\bar\phi}{}_i{}_\alpha{}^\beta, p^{(y)}_\gamma\}^{**}=
-\frac{\delta_{\alpha\gamma} + \delta_{\beta\gamma}}{2(y_\alpha+y_\beta)}\,\underset{{}^{{}^\sim}}{\bar\phi}{}_i{}_\alpha{}^\beta\,,
\end{equation}
\begin{equation}\label{CDB-2-4ap}
\{\underset{{}^{{}^\sim}}{\varphi}{}^i{}_\alpha{}^\beta, \underset{{}^{{}^\sim}}{\bar\varphi}{}_k{}_\gamma{}^\delta\}^{**}=
-i\,\frac{\delta^i_k\, \delta^\delta_\alpha \, \delta^\beta_\gamma}{y_\alpha+y_\beta} \,,\qquad
\{\underset{{}^{{}^\sim}}{\phi}{}^i{}_\alpha{}^\beta, \underset{{}^{{}^\sim}}{\bar\phi}{}_k{}_\gamma{}^\delta\}^{**}=
-i\,\frac{\delta^i_k\, \delta^\delta_\alpha \, \delta^\beta_\gamma}{y_\alpha+y_\beta}\,.\qquad
\end{equation}
Despite this complication, the calculation of the supercharge algebra  in the ``undertilded''
variables \p{CDB-2-1ap}-\p{CDB-2-4ap} is simpler than in the original ``untilded'' ones \p{CDB-2-nonh}.


\begin{thebibliography}{99}




\bibitem{AP}
V.\,Akulov, A.\,Pashnev, {\it Quantum superconformal model in $(1,2)$ space}, Teor.~Mat.~Fiz. {\bf 56} (1983) 344.

\bibitem{FR}
S.\,Fubini, E.\,Rabinovici, {\it Superconformal quantum mechanics}, Nucl. Phys. B {\bf 245} (1984) 17.

\bibitem{IKL2}
E.\,Ivanov, S.\,Krivonos, V.\,Leviant,
{\it Geometric superfield approach to superconformal mechanics},
J.\,Phys. A: Math. Gen. {\bf 22} (1989) 4201.


\bibitem{7}
J.A.\,de Azc\'arraga, J.M.\,Izquierdo, J.C.\,P\'erez Bueno,
P.K.\,Townsend, {\it Superconfor\-mal mechanics and nonlinear realizations},
Phys. Rev. D {\bf 59} (1999) 084015, {\tt arXiv:hep-th/9810230}.


\bibitem{nscm}
P.\,Claus, M.\,Derix, R.\,Kallosh, J.\,Kumar, P.K.\,Townsend,
A.\,Van Proeyen, {\it Black holes and superconformal mechanics}, Phys. Rev. Lett. {\bf 81} (1998) 4553,  {\tt arXiv:hep-th/9804177}.


\bibitem{IL}
E.\,Ivanov, O.\,Lechtenfeld,
{\it ${\mathcal{N}}{=}\,4$ Supersymmetric Mechanics in Harmonic Superspace},
JHEP {\bf 0309} (2003) 073, {\tt arXiv:hep-th/0307111}.


\bibitem{ikl}
E.\,Ivanov, S.\,Krivonos, O.\,Lechtenfeld, {\it New variant of ${\cal N}{=}\,4$ superconformal mechanics}, JHEP {\bf 0303} (2003) 014,
{\tt arXiv:hep-th/0212303}.

\bibitem{IvKrN}
E.\,Ivanov, S.\,Krivonos, J.\,Niederle, {\it Conformal and
superconformal mechanics revisited}, Nucl. Phys. B {\bf 677} (2004)
485, {\tt arXiv:hep-th/0210196}.

\bibitem{ikl1}
E.\,Ivanov, S.\,Krivonos, O.\,Lechtenfeld, {\it ${\cal N}{=}\,4$, $d{=}\,1$ supermultiplets from nonlinear realizations of $D(2,1;\alpha)$},
Class.~Quant. Grav. {\bf 21} (2004) 1031, {\tt arXiv:hep-th/0310299}.


\bibitem{superc}
S.\,Fedoruk, E.\,Ivanov, O.\,Lechtenfeld,
{\it Superconformal mechanics},
J. Phys. A  {\bf 45} (2012) 173001,
{\tt arXiv:1112.1947 [hep-th]}.


\bibitem{BIKL}
S.\,Bellucci, E.\,Ivanov, S.\,Krivonos, O.\,Lechtenfeld, {\it ${\cal N}{=}\,8$ superconformal mechanics},
Nucl. Phys. B {\bf 684} (2004) 321, {\tt arXiv:hep-th/0312322}.

\bibitem{F4scm}
F.\,Delduc, E.\,Ivanov,
{\it New model of ${\cal N}{=}\,8$ superconformal mechanics},
Phys. Lett. B {\bf 654} (2007) 200, {\tt arXiv:0706.2472\,[hep-th]}.

\bibitem{AzKuTo}
N.\,Aizawa, Z.\,Kuznetsova, F.\,Toppan,
{\it The quasi-nonassociative exceptional $F(4)$ deformed quantum oscillator},
J. Math. Phys. {\bf 59} (2018) 022101, {\tt arXiv:1711.02923\,[math-ph]}.

\bibitem{KLS-18}
S.\,Krivonos, O.\,Lechtenfeld, A.\,Sutulin,
{\it  $\mathcal{N}$-extended supersymmetric Calogero models},
Phys. Lett. B {\bf 784} (2018) 137,
{\tt arXiv:1804.10825 [hep-th]}.


\bibitem{KuTo}
Z.\,Kuznetsova, F.\,Toppan,
{\it D-module representations of ${\cal N}{=}\,2,4,8$ superconformal algebras and
their superconformal mechanics},  J. Math. Phys. {\bf 53} (2012) 043513, {\tt arXiv:1112.0995\,[hep-th]};

\bibitem{KhoTo}
S.\,Khodaee, F.\,Toppan,
{\it Critical scaling dimension of D-module representations of ${\cal N}{=}\,4,7,8$ Superconformal Algebras and constraints on Superconformal Mechanics},
J. Math. Phys. {\bf 53} (2012) 103518, {\tt arXiv:1208.3612\,[hep-th]}.

\bibitem{F4-r}
U.\,Gran, J.\,Gutowski, G.\,Papadopoulos,
{\it  Classification, geometry and applications of supersymmetric backgrounds},
{\tt arXiv:1808.07879 [hep-th]}.


\bibitem{F4-n}
S.\,Beck, U.\,Gran, J.\,Gutowski, G.\,Papadopoulos,
{\it  All Killing Superalgebras for Warped AdS Backgrounds},
{\tt arXiv:1710.03713 [hep-th]}.

\bibitem{F4-l}
G.\,Dibitetto, G.\,Lo\,Monaco, A.\,Passias, N.\,Petri, A.\,Tomasiello,
{\it  $AdS_3$ solutions with exceptional supersymmetry},
{\tt arXiv:1807.06602 [hep-th]}.



\bibitem{FrLin-1}
E.S.\,Fradkin, V.Ya.\,Linetsky,
{\it  An Exceptional $\mathcal{N}{=}\,8$ superconformal algebra in two-dimensions associated with $F(4)$},
Phys. Lett. B {\bf 275} (1992) 345.

\bibitem{FrLin-2}
E.S.\,Fradkin, V.Ya.\,Linetsky,
{\it  Results of the classification of superconformal algebras in two-dimensions},
Phys. Lett. B {\bf 282} (1992) 352, {\tt arXiv:hep-th/9203045}.


\bibitem{Poly-gauge}
A.P.\,Polychronakos,
{\it Integrable systems from gauged matrix models}, Phys. Lett.  B {\bf 266} (1991) 29-34.

\bibitem{Poly2001}
A.P.\,Polychronakos,
{\it Quantum Hall states as matrix Chern-Simons theory},
JHEP {\bf 0104} (2001) 011, {\tt arXiv:hep-th/0103013}.

\bibitem{Poly-rev}
A.P.\,Polychronakos,
{\it Physics and mathematics of Calogero particles},
J. Phys.  A {\bf39} (2006) 12793, {\tt arXiv:hep-th/0607033}.

\bibitem{FIL}
S.\,Fedoruk, E.\,Ivanov, O.\,Lechtenfeld,
{\it Supersymmetric Calogero models by gauging},
Phys. Rev. D {\bf 79} (2009) 105015,
{\tt arXiv:0812.4276\,[hep-th]}.

\bibitem{FI}
S.\,Fedoruk, E.\,Ivanov,
{\it Gauged spinning models with deformed supersymmetry},
JHEP {\bf 1611} (2016) 103,
{\tt arXiv:1610.04202\,[hep-th]}.

\bibitem{FILS}
S.\,Fedoruk, E.\,Ivanov, O.\,Lechtenfeld, S.\,Sidorov,
{\it Quantum $SU(2|1)$ supersymmetric Calogero-Moser spinning systems},
JHEP {\bf 1804} (2018) 043,
{\tt arXiv:1801.00206\,[hep-th]}.

\bibitem{2}
F.\,Delduc, E.\,Ivanov, {\it Gauging ${\cal N}{=}\,4$ Supersymmetric Mechanics}, Nucl. Phys. B {\bf 753} (2006) 211,
{\tt arXiv:hep-th/0605211};
{\it Gauging ${\cal N}{=}\,4$ supersymmetric mechanics II: (1,4,3) models from the (4,4,0) ones},
Nucl. Phys. B {\bf 770} (2007) 179, {\tt arXiv:hep-th/0611247}.

\bibitem{Bykov}
D.V.\,Bykov, {\it Classical solutions of a flag manifold $\sigma$-model},  Nucl. Phys. B {\bf 902} (2016) 292-301,
{\tt  arXiv:1506.08156 [hep-th]}.


\bibitem{TS-2018}
Y.\,Tanizaki, T.\,Sulejmanpasic,
{\it Anomaly and global inconsistency matching: $\theta$-angles, $\mathrm{SU}(3)/\mathrm{U}(1)^2$ nonlinear sigma model,
$\mathrm{SU}(3)$ chains and its generalizations},
Phys. Rev. B {\bf 98} (2018) 115126, {\tt arXiv:1805.11423 [cond-mat.str-el]}.

\bibitem{OhSeSh}
K.\,Ohmori, N.\,Seiberg, S.-H.\,Shao, {\it Sigma Models on Flags}, {\tt arXiv:1809.10604 [hep-th]}.

\bibitem{GIKOS-N3}
A.\,Galperin, E.\,Ivanov, S.\,Kalitzin, V.\,Ogievetsky, E.\,Sokatchev,
{\it  Unconstrained off-shell $\mathcal{N}{=}\,3$ supersymmetric Yang-Mills theory},
Class. Quant. Grav. {\bf 2} (1985) 155.

\bibitem{IST}
E.\,Ivanov, S.\,Sidorov, F.\,Toppan,  {\it Superconformal mechanics in $SU(2|1)$ superspace},\\
Phys. Rev. D {\bf 91} (2015) 085032, {\tt arXiv:1501.05622 [hep-th]}.

\bibitem{ILS1}
E.\,Ivanov, O.\,Lechtenfeld, S.\,Sidorov, {\it $SU(2|2)$ supersymmetric mechanics}, \\
JHEP {\bf 1611}  (2016) 031, {\tt arXiv:1609.00490 [hep-th]}.

\bibitem{ILS2}
E.\,Ivanov, O.\,Lechtenfeld, S.\,Sidorov, {\it Deformed ${\cal N}{=}\,8$ mechanics of (8,8,0) multiplets},\\
JHEP {\bf 1808}  (2018) 193, {\tt arXiv:1807.11804[hep-th]}.

\bibitem{FI-2015}
S.\,Fedoruk, E.\,Ivanov,
{\it New realizations of the supergroup D(2,1;$\alpha$) in $\mathcal{N}{=}\,4$ superconformal mechanics},
JHEP {\bf 1510} (2015) 087, {\tt arXiv:1507.08584\,[hep-th]}.



\bibitem{leva}
E.\,Ivanov, S.\,Krivonos, V.\,Leviant,
{\it Geometric superfield approach to superconformal mechanics},
J.\,Phys.\,A: Math. Gen. {\bf 22} (1989) 4201.

\bibitem{ABC}
S.\,Bellucci, E.\,Ivanov, S.\,Krivonos, O.\,Lechtenfeld,
{ABC of $\mathcal{N}{=}\,8$, $d{=}\,1$ supermultiplets},
Nucl. Phys. B {\bf 699} (2004) 226, {\tt arXiv:hep-th/0406015}.


\bibitem{FRS}
L.\,Frappat, P.\,Sorba, A.\,Sciarrino,
{\it Dictionary on Lie superalgebras}, Academic Press 2000, 410 p.,
{\tt arXiv:hep-th/9607161}.

\bibitem{VP}
A.\,Van Proeyen, {\it Tools for supersymmetry},
{\tt arXiv:hep-th/9910030}.



\bibitem{HornJons}
R.A.\,Horn, C.R.\,Johnson, {\it  Matrix analysis}, Cambridge University Press 2013, 2nd ed., 643 p.


\bibitem{Kallosh-1985}
R.E.\,Kallosh,
{\it  Superstrings and harmonic superspace},
in ``Quantum field theory and quantum statistics'', Vol. 2, Essays in honor of the 60th birthday of E.S.\,Fradkin,
eds. I.A.\,Batalin et al., 485-505.


\bibitem{Bandos-1988}
I.A.\,Bandos,
{\it  Solution of linear equations in spaces of harmonic variables},
Theor. Math. Phys. {\bf 76} (1988) 783.

\bibitem{GIKOS}
A.S.\,Galperin, E.A.\,Ivanov, S.\,Kalitzin, V.I.\,Ogievetsky, E.S.\,Sokatchev, {\it Unconstrained ${\cal N}{=}\,2$ matter, Yang-Mills and supergravity
theories in harmonic superspace}, Class. Quant. Grav.
{\bf 1} (1984) 469.

\bibitem{HSS}
A.S.\,Galperin, E.A.\,Ivanov, V.I.\,Ogievetsky and E.S.\,Sokatchev,
{\it Harmonic Superspace}, Cambridge University Press 2001, 306 p.

\bibitem{Gal-2017}
A.\,Galajinsky, {\it Couplings in $D(2, 1;\alpha)$ superconformal mechanics from the $\mathrm{SU}(2)$ perspective},
JHEP {\bf 1703} (2017) 054, {\tt arXiv:1702.01955 [hep-th]}.


\bibitem{Gal-2015}
A.\,Galajinsky, {\it $\mathcal{N}{=}\,4$ superconformal mechanics from the $su(2)$ perspective},
JHEP {\bf 1502} (2015) 091, {\tt arXiv:1412.4467 [hep-th]}.


\bibitem{BMW-1982}
G.\,Vanden\,Berghe, H.\,De\,Meyer, P.\,De\,Wilde,
{\it  $SO(2n+1)$ in an $SO(2n-3)\otimes SU(2)\otimes SU(2)$ basis. II. Detailed study of the symmetric representations of the $SO(7)$ group},
J.~Phys. A {\bf 15} (1982) 2677.

\bibitem{JW-1984}
J.\,Van\,der\,Jeugt, P.\,De\,Wilde,
{\it  The shift operator technique for $SO(7)$ in an $[SU(2)]^3$ basis. I. Theory},
J.~Math.~Phys. A {\bf 25} (1984) 2953.


\bibitem{GLP-2007}
A.\,Galajinsky, O.\,Lechtenfeld, K.\,Polovnikov,
{\it $\mathcal{N}{=}\,4$ superconformal Calogero models},
JHEP {\bf 0711} (2007) 008, {\tt arXiv:0708.1075 [hep-th]}.



\end{thebibliography}
\end{document}